\documentclass[reprint,prd,amsmath,showpacs,floatfix]{revtex4-1}
\pdfoutput=1
\usepackage[pdftex]{graphicx}
\usepackage{hyperref}
\usepackage{bm}
\usepackage{natbib}
\usepackage{ulem}
\usepackage{epstopdf}
\usepackage{braket}
\usepackage{float}
\usepackage[caption=false]{subfig}
\usepackage{amssymb}

\newcommand{\mm}{\mu\bar{\mu}}
\newcommand{\ee}{e\bar{e}}
\newcommand{\kp}{k_{\perp}}
 \begin{document} \title{True Muonium $(\mu^+ \mu^-)$ on the Light
 Front}

\author{Henry Lamm}
\email{hlammiv@asu.edu}
\author{Richard F. Lebed}
\email{Richard.Lebed@asu.edu}
\affiliation{Department of Physics, Arizona State University, Tempe,
AZ 85287}
\date{April, 2014}

\begin{abstract}
  Applying Discretized Light Cone Quantization, we perform the first
  calculation of the spectrum of true muonium, the $\mu^+ \mu^-$ atom,
  as modified by the inclusion of an $|\ee\rangle$ Fock component.
  The shift in the mass eigenvalue is found to be largest for triplet
  states.  If $m_e$ is taken to be a substantial fraction of $m_\mu$,
  the integrated probability of the electronic component of the $1 \,
  {}^3 \! S_1$ state is found to be as large as $O(10^{-2})$.  Initial
  studies of the Lamb shift for the atom are performed.  Directions
  for making the simulations fully realistic are discussed.
\end{abstract}

\maketitle
\section{Introduction}
\label{sec:intro}
The atom consisting of a $\mu^+ \mu^-$ bound state, called {\it true
  muonium\/} [in distinction from conventional muonium, the atom
($\mu^+ e^-$)] has not yet been identified, despite much more exotic
bound states having been successfully characterized, such as $\pi
\mu$ atoms~\cite{Coombes:1976hi} and even the dipositronium ($e^+
e^-$)($e^+ e^-$) molecule~\cite{Cassidy:2007}.  While ($\mu^+ \mu^-$)
atoms have doubtless been produced many times, for example in $e^+
e^-$ colliders (where they can be easily lost in the beam line before
decaying), it is only recently that immediately realizable experiments
have been proposed to generate sufficient numbers of true muonium
atoms to guarantee their unambiguous observation.  Ideas include $e^+
e^-$ collider experiments in which the beams intersect at an angle
rather than head on [thus projecting ($\mu^+ \mu^-$) along the beam
bisector] or conventional setups in which the produced ($\mu^+ \mu^-$)
recoils against a co-produced photon~\cite{Brodsky:2009gx}, or
producing ($\mu^+ \mu^-$) as a byproduct of experiments that scatter
electron beams from high-$Z$ fixed targets in search of hidden-sector
{\it heavy photons\/}~\cite{Banburski:2012tk}.

True muonium consists of a spectrum of metastable states (lifetimes in
the ps to ns range~\cite{Brodsky:2009gx}), for which the 2.2~$\mu$s
muon weak decay lifetime is effectively infinite.  The transitions are
therefore overwhelmingly electromagnetic, meaning that ($\mu^+ \mu^-$)
can be treated, like positronium $(e^+ e^-)$, as a Bohr atom [in the
($\mu^+ \mu^-$) case, with a ground-state radius of only 512~fm], and
its transitions can be computed using QED.\@ Unlike positronium, its
decay products include not only monochromatic photons, but also $e^+
e^-$ pairs of well-characterized energies.  Studies of ($\mu^+ \mu^-$)
are further motivated by discrepancies in precision muon-related
effects, such as $(g - 2)_\mu$ and the proton charge
radius~\cite{Antognini:1900ns,TuckerSmith:2010ra}.

As far as an isolated ($\mu^+ \mu^-$) bound state is concerned,
conventional QED and quantum-mechanical wave functions are generally
sufficient to compute its most important features, just as is true for
($e^+ e^-$).  However, these techniques provide not as much guidance
for understanding the detailed structure-dependent dynamics of the
atoms.  Indeed, remarkable experiments can in principle be performed
on these ``atoms in flight'', including measurements of the Lamb
shift~\cite{Robiscoe:1965}.  Bound states in quantum field theory
(QFT) are notoriously complicated objects, due not only to the fact
that the vacuum consists of an arbitrarily large number of virtual
particles, but also because the constituents can be relativistic, and
the required boosts to relate the components of the wave function
depend in a detailed way upon the interaction.

In fact, a well-known method of studying bound states avoids these
problems.  By performing quantization not, as usual, with respect to
conventional time $t$ (called {\it instant form}) but rather with
respect to the light-front time $x^+ \equiv t + z$ (called {\it front
  form})~\cite{Dirac:1949cp}, one develops a Hamiltonian formalism
that is nonetheless fully covariant and ideally suited to
characterizing bound states comprised of well-defined
constituents~\cite{Brodsky:1997de}.  The analogue to a Schr\"{o}dinger
equation for the state then becomes an infinite but denumerable set of
coupled integral equations.  In order to obtain numerical results, one
may then truncate the equations by limiting the set of component Fock
states included in the calculation and discretizing momenta with
suitable periodic boundary conditions, effectively turning
infinite-dimensional matrices into finite ones and creating a problem
solvable on a computer, which is termed Discretized Light-Cone
Quantization (DLCQ)~\cite{Pauli:1985ps}.

Our purpose in this article is to report on the first simulation of
the true muonium atom using DLCQ and other light-front techniques.  We
borrow heavily on older works that use this approach to simulate
positronium.  The first such attempt~\cite{Tang:1991rc} was impeded by
slow numerical convergence, which was successfully addressed by a
better treatment~\cite{Krautgartner:1991xz} of the singularity induced
by the Coulomb interaction.  Further work~\cite{Kaluza:1991kx}
superseded the effective interaction approach
of~\cite{Krautgartner:1991xz} to perform a direct diagonalization of
the Hamiltonian matrix.  The work of Trittmann and
Pauli~\cite{Trittmann:1997xz} unified these improvements, and
additionally included in the Fock space the explicit single-photon
annihilation channel $|\gamma \rangle$.  They also demonstrated the
numerical restoration of rotational invariance, a needed improvement
since light-front coordinates treat the longitudinal and transverse
directions very differently.  Our work is a direct application of the
methods introduced in Ref.~\cite{Trittmann:1997xz} to true muonium;
our principal innovation, apart from the trivial substitution of the
valence electrons with muons, is the inclusion of $|\ee \rangle$ Fock
states, which can mix with the $|\mm \rangle$ component via the
explicit $|\gamma\rangle$ component.  We also apply a subtraction to
the interaction amplitudes in order to regularize the ultraviolet
behavior of observables.

The positronium simulations of Ref.~\cite{Trittmann:1997xz} adopted
the unphysically large value $\alpha = 0.3$ in order to show the
robustness of the method at large values of coupling, where high-order
QED calculations begin to become unreliable.  An additional purpose
for this choice was to provide numerical evidence that strong-coupling
DLCQ would be applicable to the much more intricate QCD bound-state
problem.  In our calculations presented here, we maintain this large
value of $\alpha$ in order to study the effects of a large $|\ee
\rangle$ component of true muonium; of course, $\alpha$ is an
adjustable parameter of the program and can be altered for more
physical simulations in our future work.  Moreover, the code designed
by Ref.~\cite{Trittmann:1997xz} contains only one fermion mass
parameter, $m_e$ in their case and $m_\mu$ in ours.  In the present
work, $m_e$ is an independent parameter; again, to study the effects
of a large $|\ee \rangle$ component, we allow $m_e$ to be any finite
fraction of $m_\mu$.  Ultimately, for physical applications one sets
$m_e/m_\mu \simeq 4.8\cdot 10^{-3}$, but we exhibit results with
$m_e/m_\mu = O(1)$ for the practical reason that properly
characterizing a system in QFT with two widely separated scales
requires proper renormalization evolution between the scales (In front
form, the original diagrammatic renormalization techniques are
described in Ref.~\cite{Brodsky:1973kb}).  Since the evolution
typically depends upon the log of the ratio of scales, choosing $m_e$
not excessively small compared to $m_\mu$ minimizes this effect; we
plan to address this important issue thoroughly in a subsequent
paper~\cite{LamLeb2}.

Even with a large $|\ee \rangle$ component, our simulations reproduce
the Bohr spectrum of true muonium to a high degree of precision.
Nevertheless, the significance of the $|\ee \rangle$ component is
clearly seen in the true muonium eigenvalues, and especially in the
wave functions.  We also see modifications to the $(\mu^+ \mu^-)$ Lamb
shift in agreement with its expected dependence upon the atom's
principal quantum number $n$.  While the proper numerical and formal
treatment of the $|\ee \rangle$ sector will require future
improvements (the effects of neglecting which are exhibited and
discussed in detail below), we find these initial results to be very
encouraging.

This paper is organized as follows.  In Sec.~\ref{sec:lightfront}, we
present essential conceptual and formal details of light-front
bound-state techniques relevant to this work, and describe the
associated numerical details in Sec.~\ref{sec:num}\@.  We define our
model in Sec.~\ref{sec:model} and discuss the effects of the
$|\ee\rangle$ component in Sec.~\ref{sec:ee}\@.
Section~\ref{sec:cutoffs} provides a discussion of the effect of
momentum-space cutoffs, lepton mass renormalization, and the
regularization of interaction amplitudes.  We present our results in
Sec.~\ref{sec:results}, with conclusions in Sec.~\ref{sec:con} and
details of the calculation of the relevant matrix elements in the
Appendix.

\section{Light-Front Bound States} \label{sec:lightfront}

The action-functional approach has become the predominant method for
performing QFT calculations.  It is based upon a Lorentz-invariant
Lagrangian density, and the symmetries of the theory (including gauge
invariance) can be incorporated very efficiently.  On the other hand,
the Hamiltonian formalism provides a more natural language for
describing bound-state systems, just as it does in non-relativistic
quantum mechanics.  However, Hamiltonians are complicated objects in
relativistic QFT, especially in the instant form: The nonanalytic
nature of the operator $\sqrt{ \bm{P}^2 + M^2 }$ causes difficulties;
the constituent particles have distinct rest frames, and the boost
operations necessary to express them as part of a single wave function
depend intrinsically upon the interactions between them; and most
significantly, the vacuum structure is exceedingly complicated due to
the creation and annihilation of virtual states with arbitrarily large
numbers of quanta.

The front form addresses each of these difficulties.  First, the
Hamiltonian operator $P^- \equiv P^0 - P^3$ conjugate to the time
coordinate $x^+$ is related linearly to the energy-squared operator,
\begin{equation}
  H_{\rm LC} = P^- P^+ - \bm{P}^2_\perp \, ,
\end{equation}
where $P^+ \equiv P^0 + P^3$ is the longitudinal momentum component
conjugate to the light-front longitudinal direction $x^- \equiv t -
z$, and the transverse momentum is $\bm{P}_\perp \equiv P^1 \bm{x} +
P^2 \bm{y}$.  For an eigenvalue $M^2$ of the operator $H_{\rm LC}$
corresponding to a state $| \Psi \rangle$, the light-front Hamiltonian
eigenvalue equation is no longer nonanalytic:
\begin{equation}
P^- \left| \Psi \right> = \frac{M^2 + \bm{P}_\perp^2}{P^+} \left|
\Psi \right> \, .
\label{eq:LFHam1}
\end{equation}

Second, out of the 10 Poincar\'{e} generators $P^\mu$ and $M^{\mu
\nu}$, the front form is special in having 7 that leave the time slice
$x^+ =$~{\it const\/} invariant~\cite{Dirac:1949cp}, and therefore can
be expressed without reference to the specific interaction in the
Hamiltonian.  These so-called {\it kinematic\/} operators are the
spatial momentum components $\bm{P}_\perp$, $P^+$, the longitudinal
angular momentum ($M^{12} = J_z$) and boost ($M^{+-} = K_z$)
generators, and the mixed transverse-longitudinal boosts $M^{+
\perp}$.  In contrast, the instant form has only 6 kinematic operators:
the spatial momentum $P^i$ and angular momentum $M^{ij}$ components;
in particular, the boost generators $M^{0i}$ are all {\it dynamic\/},
in that they depend upon the interaction.  Although the front form
addresses the boost issue by providing kinematic boost generators, the
price is the loss of manifest rotational invariance (the absence of
simple $\bm{J}_\perp$ generators), which is apparent from the fact
that front form treats $x^3$ differently from $x^{1,2}$.  The $P^\mu$
and total spin operators $S^2$ and $S_z$ still commute in front form,
meaning that a particular state $\left| \Psi \right>$ can be
completely specified as
\begin{equation}
\left| \Psi ; M, P^+ \! , \bm{P}_\perp , S^2 , S_z ; h \right> \, ,
\end{equation}
where $h$ indicates any discrete or non-spacetime quantum numbers,
such as parity or lepton number.

Third, and most significantly, the numerical values of longitudinal
momenta $P^+$ for all physical particles are nonnegative, since $P^0 =
E > P^3$.  One cannot create virtual particles traveling ``backwards''
with respect to the light-front longitudinal direction, so empty space
cannot produce collections of virtual particles in front form, in
stark contrast to the situation in instant form.  The ground state of
the free theory is also the ground state of the full interacting
theory, and the Fock state expansion built upon the free vacuum
provides a rigorous ``parton'' component description of the full
interacting state.

To be specific, the state $\left| \Psi \right>$ may be expressed in
terms of its Fock components $\left| \mu_n \right>$, where $n$ in
general is denumerably infinite.  Each particular component $\left|
  \mu_n \right>$ contains a fixed number $N_n$ of constituent quanta,
the $i^{\rm th}$ of which has rest mass $m_i$ and momentum $k^\mu_i$
(out of the total momentum $P^\mu$).  The kinematics may alternatively
be described in terms of longitudinal boost-invariant quantities $x_i
\equiv k^+_i/P^+$ ($0 \leq x_i \leq 1$) and $\bm{k}_{\perp i}$, and
helicities $\lambda_i$, so that
\begin{equation}
\sum_{i=1}^{N_n} x_i = 1, \ \ \sum_{i=1}^{N_n} \bm{k}_{\perp i} =
\bm{P}_\perp \, ,
\label{eq:kineconstraint}
\end{equation}
and, working in the {\it intrinsic frame\/}, in which $\bm{P}_\perp =
0$,
\begin{equation}
k_i^\mu = \left( x_i P^+ \! \! , \, \bm{k}_{\perp i} , \frac{m_i^2 +
\bm{k}_{\perp i}^2} {x_i P^+} \right) \, .
\end{equation}
Using the completeness of the states $\left| \mu_n \right>$, the
decomposition then reads
\begin{eqnarray}
\left| \Psi \right> & \equiv & \sum_n \int
[{\rm d}\mu_n] \left| \mu_n^{\vphantom{x}} \right> \left< \mu_n
\right| \left. \! \Psi ; M , P^+ \! , \bm{P}_\perp , S^2 , S_z ; h
\right> \nonumber \\ & \equiv & \sum_n \int [{\rm d}\mu_n] \left|
\mu_n \right> \Psi_{n|h} (\mu) \, ,
\end{eqnarray}
where the measure notation indicates an integration over values of all
constituent $x_i$, $\bm{k}_\perp$, subject to the constraints of
Eq.~(\ref{eq:kineconstraint}).  The functions $\Psi_{n|h} (\mu)$,
where now $h$ and $\mu$ are shorthand for all the intrinsic and
kinematic quantum numbers, respectively, of the Fock state $n$, are
called the {\it component wave functions\/} of the state and are the
central objects of interest in bound-state light-front calculations.
The Hamiltonian expression Eq.~(\ref{eq:LFHam1}) then becomes
\begin{eqnarray}
\lefteqn{\sum_{n'} \int [{\rm d}\mu_{n'}] \left< \mu_{n} \! :  x_i ,
\bm{k}_{\perp i} , \lambda_i \right| P^- \! \left| \mu_{n'} \! :
x'_i , \bm{k}'_{\perp i} , \lambda^\prime_i\right>} \hspace{8em} & &
\nonumber \\ & & \hspace{3.3em} \times \Psi_{n' | h} (x'_i,
\bm{k}'_{\perp i} , \lambda^\prime_i ) \nonumber
\\ & = & \frac{M^2 + \bm{P}_\perp^2}{P^+} \Psi_{n | h}
(x_i, \bm{k}_{\perp i} , \lambda_i ) \, ,
\label{eq:LFHamexact}
\end{eqnarray}
which is an exact infinite-dimensional integral equation for the
component wave functions $\Psi_{n|h} (\mu)$.  Although this expression
has been derived from a full QFT with no approximations, one may
identify it as the light-front version of the Schr\"odinger equation.
Specifically, the kinetic energy operator for the $i^{\rm th}$
component in the frame $\bm{P}_\perp = 0$ reads
\begin{equation}
T_i = \frac{ m_i^2 + \bm{k}_{\perp i}^2 }{x_i} \, ,
\end{equation}
and the bound-state equation for two equal-mass valence particles
interacting via the effective potential $V_{\rm eff}$ is described by
the integral equation
\begin{eqnarray}
\label{eq:LFHam2}
\lefteqn{\left(M^2 - \frac{m^2+\bm{k}^{2}_{\perp}}{x(1-x)} \right)
\psi (x,\bm{k}_\perp;\lambda_1,\lambda_2 )} \nonumber \\
& = &\sum_{\lambda'_1,\lambda'_2}\int_D
\mathrm{d}x' \mathrm{d}^2 \bm{k}'_\perp
\langle x,\bm{k}_\perp;\lambda_1,\lambda_2 \left| V_{\rm eff}
\right| x, \bm{k}_\perp;\lambda'_1,\lambda'_2, \rangle \nonumber\\ & &
\hspace{2em} \times \psi (x',\bm{k}'_\perp; \lambda'_1,\lambda'_2 ) \, .
\end{eqnarray}
The appearance of reciprocal powers of momenta in
Eq.~(\ref{eq:LFHam1}), which is the ultimate origin of the
singularities at $x = 0$ or 1 in Eq.~(\ref{eq:LFHam2}), requires a
careful regularization of numerical integrals.  The domain $D$ in
Eq.~(\ref{eq:LFHam2}) is defined by introducing a cutoff $\Lambda$ on
the parton transverse momentum $\bm{k}_\perp$; in the equal-mass case,
we choose~\cite{Lepage:1980fj}
\begin{equation}
\frac{ m^2 + \bm{k}_\perp^2 }{x(1-x)} \le \Lambda^2 + 4m^2 \, .
\end{equation}
Instituting a momentum-space cutoff has the added effect of minimizing
the influence of multiparticle Fock states.  In principle, each sector
of the theory (in our case, notably $\mm$ and $\ee$) can have an
independent cutoff, but such choices must be motivated by the physical
scales of the problem, and in any case the final results must
eventually be insensitive to such particular choices.  We discuss
these issues in greater detail in Sec.~\ref{sec:cutoffs}.

For a gauge theory like QED, the light-cone gauge $A^+ = 0$
is the most natural choice because it eliminates the spatial
non-transverse modes.  Using the equations of motion, one may then
eliminate the $A^-$ component in favor of the other fields in the
theory; this inversion is subtle due to the existence of ``zero
modes'' of $A^+$, but such modes are not expected to affect the
spectrum of bound states in a crucial way~\cite{Trittmann:1997xz}.
The result is a fairly complicated but closed-form exact Hamiltonian
that may be used to develop front-form Feynman
rules~\cite{Brodsky:1997de,Lepage:1980fj}.

The Hamiltonian can be described by the sum of Feynman rules for a
kinetic operator $T$ and various types of interactions:
\textit{seagulls} $S$ [including their normal-ordered
\textit{contractions}], which do not change particle number,
\textit{vertices} $V$, which change particle number by one, and
\textit{forks} $F$, which change particle number by two:
\begin{equation}
 H_{\rm LC}=T+S+V+F \, .
\end{equation}
The exact form of each operator has been worked out and can be found
in many places, {\it e.g.}, in Ref.~\cite{Brodsky:1997de}.  The
connection of the lowest Fock states by these interactions for true
muonium is summarized in Table~\ref{tab:hamil}.
\def\d{$\bullet$} \def\V{ V } \def\b{ $\cdot$ } \def\s{ S } \def\f{ F }
\begin{table}[ht]
\centerline{
\begin {tabular}{|l|r||ccc|cccccc|}
\hline 
  \rule[-3mm]{0mm}{8mm}  Sector & $n$ & 
     0 & 1 & 2 & 3 & 4 & 5 & 6 & 7 & 8
\\ \hline \hline   
    $ |\gamma \rangle$ &  0 & 
    \d & \V &\V &\b &\f &\f &\b &\b &\b  
\\
    $ |e\bar e\rangle $ &  1 & 
    \V &\d &\s &\s &\V &\b &\f &\b &\f
\\
    $ |\mu\bar \mu\rangle $ &  2 & 
    \V &\s &\d &\s &\b &\V &\b &\f &\f
    \\
    \hline
    $ |\gamma \gamma\rangle$ &  3 & 
    \b & \s &\s &\d &\V &\V &\b &\b &\b
\\
    $ |e\bar e \, \gamma \rangle$ &  4 & 
    \f &\V &\b&\V &\d &\s &\V &\b &\V
\\
    $ |\mu\bar \mu \, \gamma \rangle$ &  5 & 
    \f &\b &\V &\V &\s &\d &\b &\V &\V
\\
    $ |e\bar e\, e\bar e\rangle $ &  6 & 
    \b&\f &\b &\b &\V &\b &\d &\b &\s
\\ 
    $ |\mu\bar \mu\, \mu\bar \mu\rangle $ &  7 & 
    \b&\b &\f &\b &\b &\V &\b &\d &\s
\\
    $ |\mu\bar \mu\, e\bar e\rangle $ &  8 & 
    \b&\f &\f &\b &\V &\V &\s &\s  &\d
\\
\hline
\end {tabular}
}
\caption{The Hamiltonian matrix for two-flavor QED, where $n$ labels
  Fock states.  The vertex, seagull and fork interactions are
  denoted by V, S, F respectively.  Diagonal matrix elements are
  indicated by \d, and vanishing matrix elements by a $\cdot$.}
\label{tab:hamil}
\end{table}

In order to reduce Eq.~(\ref{eq:LFHamexact}) to a finite-dimensional
equation that can be solved numerically [in particular, the form
Eq.~(\ref{eq:LFHam2})], the space of included Fock states $\{ |F
\rangle \}$ must be truncated.  The straightforward approach of using
the fundamental interactions of $\{ |F \rangle \}$ alone and ignoring
all others $\{ |\bar F \rangle \}$ is called the {\it Tamm-Dancoff
method\/}~\cite{Tamm:1945qv,Dancoff:1950ud}.  Unfortunately, this
approach can easily sacrifice Lorentz covariance or gauge invariance;
including the effect of higher Fock states $\{ |\bar F \rangle \}$ to
restore the necessary structures is the aim of the {\it method of
iterated resolvents\/}~\cite{Pauli:1996dm}, which treats $\{ |\bar F
\rangle \}$ as giving rise to effective interactions among the states
$\{ | F \rangle \}$ [which, in Eq.~(\ref{eq:LFHam2}), contribute to
$V_{\rm eff}$].  Since only the complete theory (QED in our case) with
arbitrarily complicated Fock states that are precisely related by the
exact Lagrangian can uniquely specify the correct effective
interaction, the iterated resolvents simply act to provide one minimal
completion of the interaction, written exclusively in terms of $\{ | F
\rangle \}$.

\section{Numerical Simulation on the Light Front} \label{sec:num}

To solve Eq.~(\ref{eq:LFHam2}), one can discretize the momentum-space
Fourier modes and solve the resulting eigenvalue problem on a finite
grid (the DLCQ method).  Discretization in ($x,\bm{k}_\perp$) space
results in a asymmetric matrix, which significantly increases the
computational effort, so one may instead use a set of variables ($\mu$,
$\theta$, $\phi$) defined by
\begin{equation} \label{eq:xdefn}
 x=\frac{1}{2}\left(1+\frac{\mu\cos \theta }
{\sqrt{m_i^2+\mu^2}}\right) \, ,
\end{equation}
\begin{equation}
 \bm{k}_\perp = \mu ( \sin \theta \cos \phi ,
\sin \theta \sin \phi , 0 ) \, .
\end{equation}
Using these variables, one may exchange $\phi$ for the discrete
quantum number $J_z$ \cite{Trittmann:1997xz} and compute using only
$\mu$, $\theta$.  The new variable $\mu$ can be considered an
off-shell momentum, due to the relation
\begin{equation} \label{eq:mudef}
 \frac{m_i^2+\bm{k}^2_\perp}{x(1-x)}=4(\mu^2+m_i^2) \, .
\end{equation}
Since these coordinates depend upon the fermion mass $m_i$, different
sets of $\mu$, $\theta$ values result from the same sets of $x$ and
$\bm{k}_\perp$ values in the two-flavor system.

To discretize, we utilize the Clenshaw-Curtis method (unlike
Ref.~\cite{Trittmann:1997xz}, which used the Gauss-Legendre method)
for $\mu$, to take advantage of its rapid convergence and reuse of
quadrature points, and the Gauss-Chebyshev method for $\theta$ to
carefully sample the endpoints.  The $\mu$ range is
$\left[0,\frac{\Lambda_i}{2}\right]$, where $\Lambda_i$ is the
momentum cutoff for flavor $i$ (see Sec.~\ref{sec:cutoffs}), and $\cos
\theta \in [-1,1]$.  To allow for values $\Lambda_i\rightarrow
\infty$, we remap the interval to one $\in [0,1]$ via a weighting
\begin{equation} \label{eq:weight}
 f(\mu)=\frac{1}{1+\mu} \, ,
\end{equation}
where $\mu$ here is expressed in units of the Bohr momentum $\alpha
m_i/2$.  With reference to Eq.~(\ref{eq:xdefn}), if one fixes the
cutoff $\Lambda_i$ and increases the number of sampling points
$N_\mu$, this mapping $f$ then places more grid points at low $\mu$ to
better sample the wave function around $x = \frac 1 2$, and high $\mu$
to better sample around $x = 0$, 1.  From this scheme, one sees that
four numbers essentially determine the quality of numerical
calculation: the number of sampling points $N_{\mu}$ and $N_{\theta}$,
and the momentum cutoffs $\Lambda_\mu$ and $\Lambda_e$.  In cases
where the number of sampling points $N_\mu$, $N_\theta$ are set equal,
we write $N_\mu=N_\theta=N$.

The $1/\bm{q}^2$ singularity introduced by the Coulomb interaction,
where $\bm{q}$ is the momentum transfer, is handled by the same {\it
counterterm method\/} as described in Ref.~\cite{Trittmann:1997xz} and
first implemented in Ref.~\cite{Krautgartner:1991xz}: The sum over
discrete matrix elements near the Coulomb singularity is performed by
subtracting from the numerator a function that reduces the overall
degree of divergence of the sum, thus making it more rapidly
convergent.  In order to obtain an identity, one must add back the
term that was subtracted; however, the corresponding expression in
this case is realized as integral rather than a sum, which can be
numerically evaluated rather efficiently.  The only necessary
modification in the two-flavor case is to use separate counterterms in
each flavor sector.
 
\section{True Muonium Model} \label{sec:model}

Our model simply uses QED with two fermion flavors, the electron $e$
and the muon $\mu$, but in which the ``electron'' is dramatically
heavier than its physical value of $m_e/m_\mu\simeq 4.8\cdot 10^{-3}$,
in order to explore the effect of the extra flavor sector beyond a
simple ``scaled-up'' version of positronium.  Nevertheless, the
condition $m_e<m_\mu$ is always imposed, and as mentioned above, we
set $\alpha = 0.3$ to enhance the effects of the extra sector.  In
this basis, one can express the state of all possible charge-zero,
lepton family-number zero wave functions defined by:
\begin{eqnarray}
  |\Psi \rangle & = & \psi_{\mm} |\mm\rangle + \psi_{\ee} |\ee\rangle +
  \psi_\gamma|\gamma \rangle \nonumber \\ & & + \psi_{\mm\gamma}
  |{\mm\gamma} \rangle + \psi_{\ee\gamma}|\ee\gamma\rangle + \cdots \,
  .
\end{eqnarray}
The full physical problem would require not only decreasing $m_e$ and
$\alpha$, but also including hadronic polarization function
corrections to the photon propagators.

In this initial model of true muonium, we extend the Fock space
considered in Refs.~\cite{Trittmann:1997xz,Trittmann:1997up,
  Trittmann:1997tt,Trittmann:2000gk} to include a second flavor of
fermions: $|\mu\bar{\mu}\rangle$, $|\mu\bar{\mu}\gamma\rangle$,
$|e\bar{e}\rangle$, $|e\bar{e}\gamma\rangle$, and $|\gamma\rangle$.
Through a proper choice of effective interactions~\cite{Pauli:1997ns},
one can truncate the Fock space at these states and neglect other Fock
states such as $|\gamma\gamma\rangle$ or $|\mu\bar{\mu} e
\bar{e}\rangle$ in a self-consistent way.  The $|e\bar{e}\rangle$
states are of particular interest because their continuum states
constitute the dominant decay mode of the ${}^3 \! S_1$ ($C = -1$)
states of true muonium.  Furthermore, its inclusion in the numerical
simulation (along with $|e\bar{e}\gamma\rangle$) is straightforward
because it requires no conceptual innovation beyond that used to study
the $|\mu\bar{\mu}\rangle$ and $|\mu\bar{\mu}\gamma\rangle$ states;
the greatest complication is that the matrices to be diagonalized
become much larger, increasing computation time.

Solving for the eigenstates of $H_{\rm LC}$ with this limited Fock
space nonetheless gives the bound states of positronium $(\ee)$ and
true muonium $(\mm)$, as well as the continuum states of $\gamma$,
$\ee$, and $\mm$, exactly (up to the effects due to the neglected
higher-order Fock states).  The output of this calculation is the wave
functions of various helicity states for $\left|\mm\right>$ and
$\left|\ee\right>$ components $\psi$ in the form of
Eq.~(\ref{eq:LFHam2}), {\it i.e.}, with the $\left|\gamma\right>$
components folded into $V_{\rm eff}$ by means of the method of
iterated resolvents.

\section{Effect of the $\left| \ee \right>$ Sector} \label{sec:ee}
Proper inclusion of the front-form Fock states $\left|\ee\right>$ and
$\left|\ee\gamma\right>$ should replicate the physics in instant form
due to the inclusion of instant-form diagrams with an $e^+ e^-$ pair
and a $\gamma$, such as vacuum polarization due to electrons in the
single-photon annihilation channel (called VP-e-A in
Ref.~\cite{Jentschura:1997tv}).  The importance of including both
$\left|\ee\right>$ and $\left|\ee\gamma\right>$ states in our
simulations is twofold.

First, the dominant decay channel for true muonium in ${}^3 \! S_1$
states is $e^+ e^-$ production, while the dominant decay channel for
${}^1 \! S_0$ ($C=+1$) states is $\gamma \gamma$.  Note that the
well-known leading-order result for the ${}^3 \!  S_1$-${}^1 \! S_0$
hyperfine splitting, $\Delta E = \frac{7}{12} m_\mu \alpha^4$ for true
muonium, has been derived analytically in front
form~\cite{Jones:1996vy,Jones:1997cb}, so that other physical effects
sensitive to small $\mu^+ \mu^-$ separation such as $\mu^+ \mu^- \to
\gamma \to e^+ e^-$ should also be considered.

While including the
$|\ee\rangle$ state into our calculations requires essentially nothing
but duplicating the $|\mm\rangle$ states as $|\ee\rangle$ and
computing one new set of matrix elements (see Appendix), properly
including a $|\gamma\gamma\rangle$ state would require computing many
new matrix elements, developing new counterterms to regularize
singular integrals, and properly renormalizing the photon mass terms
that arise on the light front.  We plan to address these issues in
future work.

Second, Jentschura {\it et al.}~\cite{Jentschura:1997tv} showed in
instant form that VP-e-A is the second-largest correction to the
hyperfine splitting in true muonium.  The only correction that is
larger in instant form arises from vertex corrections, which are
partly incorporated in front form through the inclusion of
$|\mm\gamma\rangle$ states, but can be fully treated only through
proper renormalization.  Furthermore, Ref.~\cite{Jentschura:1997tv}
finds the energy shifts arising from the $|\gamma\gamma\rangle$ states
to be several times smaller than those from either VP-e-A or vertex
corrections.  The calculations of Ref.~\cite{Jentschura:1997tv} rely
upon the asymptotic behavior of the vacuum polarization, which for
true muonium is the limit of $m_e/m_\mu \ll 1$. In our case of
$m_e/m_\mu = O(1)$, one might expect significant corrections to the
asymptotic behavior.  To find the effect of these corrections, one can
compute the exact correction due to VP-e-A without the asymptotic
approximation.  As first shown in \cite{Karplus:1952wp}, the
leading-order radiative correction to the QED particle-antiparticle
bound-state energy spectrum due to a virtual fermion loop coupling to
the electromagnetic field with amplitude $\varphi_0$ (which, in the
nonrelativistic limit, is just the wave function) is
\begin{align}
\label{eq:vp}
 \Delta E_{\rm VP}=& \, \frac{\pi\alpha}{m_i^2}
\left(1-\frac{4\alpha}{\pi}\right)
\bar{\Pi}^{R}(4m_i^2)|\varphi_0|^2\langle\bm{S}^2\rangle
+O(\alpha^6)\nonumber\\
 =& \, \frac{\alpha^4 m_i}{4n^3}\left(1-\frac{4\alpha}{\pi}\right)
\bar{\Pi}^{R}(4m_i^2)+O(\alpha^6) \, ,
\end{align}
where $m_i$ is the mass of the bound fermion, $\bar{\Pi}^{R}(q^2)$ is
the renormalized polarization function, and $\bm{S}^2$ is the total
spin Casimir operator.  In the second line we have specialized to the
$n \, {}^3 \! S_1$ state, for which $|\varphi_0|^2=m_i^3\alpha^3/8\pi
n^3$ is the nonrelativistic squared wave function at the origin, and
$\langle\bm{S}^2\rangle = 2$.  The exact form of the one-loop vacuum
polarization function is
\begin{equation}
\label{eq:pf}
\bar{\Pi}^{R}(q^2)=\frac{\alpha}{3\pi}\left[-\frac{5}{3}-
\frac{4m_f^2}{q^2}+\left(1+\frac{2m_f^2}{q^2}\right)
f(q^2)\right] \, ,
\end{equation}
where $m_f$ is the mass of the loop fermions, and the form of $f(q^2)$
depends upon whether $q^2$ is spacelike or timelike and its size
compared to $4m_f^2$.  In the region $4m_f^2<q^2$, {\it e.g.}, for
true muonium with electron-loop corrections, one finds
\begin{equation}
\label{eq:ff}
 f(q^2)=\sqrt{1-\frac{4m_f^2}{q^2}}\ln \! \left(\frac{1+\sqrt{1
-\frac{4m_f^2}{q^2}}}{1-\sqrt{1-\frac{4m_f^2}{q^2}}}\right)
-i\pi\sqrt{1-\frac{4m_f^2}{q^2}} ,
\end{equation}
where the imaginary term signals the possibility for decay.  Inserting
Eqs.~(\ref{eq:pf})--(\ref{eq:ff}) into Eq.~(\ref{eq:vp}), taking
the $m_e \to 0$ limit, and dropping the $O(\alpha^6)$ terms, we
reproduce Eq.~(26) of \cite{Jentschura:1997tv}, which we express
slightly differently as
\begin{align}
 \Delta E_{\rm VP}(n^3S_1)=\frac{m_\mu\alpha^5}{4\pi n^3}&
\left[\frac{1}{3}\ln\left(\frac{4m_\mu^2}{m_e^2}\right)-\frac{5}{9}
-\frac{i\pi}{3}\right.\nonumber\\&\phantom{xx}\left.
+O\left(\frac{m_e^2}{4m_\mu^2}
,\ \alpha\right)\right] \, . \label{eq:NREVP}
\end{align}
In typical cases considered here ($m_e/m_\mu \sim 0.1$--0.8), this
expansion predicts relative corrections to the asymptotic form of
order 10\%, so the complete formulas
[Eqs.~(\ref{eq:vp})--(\ref{eq:ff})] are retained for our numerical
results.

In order to compare our results to those of instant-form perturbation
theory calculations, we compare shifts in $M^2$:
\begin{align}
\label{eq:dm}
 \Delta M^2& \equiv M^2_{\mu\mu}-M^2_0\nonumber\\&=(2m_\mu+B
+\Delta E)^2-(2m_\mu+B)^2 \, ,
\end{align}
where $M^2_{\mu\mu}$ is the squared mass of our model true muonium
including the $|\ee\rangle$ component, while $M^2_0$ is the squared
mass neglecting the electron Fock states, $\Delta E$ is the total
binding energy due to the presence of the $|\ee\rangle$ states, and
$B$ is the remaining binding energy terms of the atom.  In
Sec.~\ref{sec:results} we examine how well taking $\Delta E = \Delta
E_{\rm VP-e-A}$, where the latter refers to the original instant-form
expression of Eq.~(\ref{eq:vp}), matches the light-front results.

\section{Cutoff and Renormalization Issues} \label{sec:cutoffs}

In front form, the most common renormalization scheme for
two-body systems is the covariant cutoff approach of Lepage and
Brodsky~\cite{Lepage:1980fj}:
\begin{equation}
\label{eq:cut}
 \frac{m^2+\bm{k}^2_{\perp}}{x(1-x)}\leq \Lambda^2+4m^2 \, .
\end{equation}
Unfortunately, in any but the simplest models, this formulation does
not properly regularize the Hamiltonian as new sectors are added.
More recent attempts to utilize other renormalization schemes include
Pauli-Villars~\cite{Chabysheva:2009vm,Chabysheva:2010vk} and
Hamiltonian flow~\cite{Gubankova:1998wj,Gubankova:1999cx} techniques,
and methods with sector-dependent
counterterms~\cite{Karmanov:2008br,Karmanov:2012aj}.  The first steps
to study positronium using the basis light-front approach have
appeared very recently~\cite{Wiecki:2014ola}.

For the purposes of this work, we take the simplest possible
renormalization scheme by defining two covariant cutoffs via
Eq.~(\ref{eq:cut}): $\Lambda_\mu$ for the muon sector and $\Lambda_e$
for the electron sector.
Since Eqs.~(\ref{eq:mudef}), (\ref{eq:cut}) identify $\Lambda$ as a
maximum off-shell momentum for the parton of mass $m$, physical
considerations lead one to expect that $\Lambda$ should be larger for
lighter components of a single bound-state system. In the true muonium
case, choices such as $\Lambda_e^2 = \Lambda_\mu^2 + 4(m_\mu^2 -
m_e^2)$ are natural, and this is the scheme adopted here.  In
particular, $\Lambda_e$ should be chosen significantly larger than
$\Lambda_\mu$, or else the phase space for $\left|e\bar{e}\right>$
continuum states contributing to true muonium is inappropriately
truncated, leading to numerical instabilities due to undersampling of
physically significant amplitudes.  This physically appropriate choice
nevertheless leads to interesting numerical issues, as discussed in
the next section.

\begin{figure}
\begin{center}
\includegraphics[width=\linewidth]{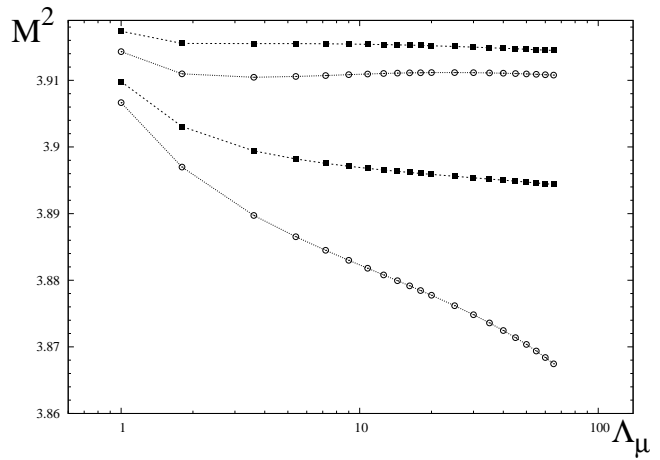}

\caption{\label{fig:cut1}Mass eigenvalues of $n=1$ true muonium states
  with $J_z=0$ ($1 \, {}^3 \! S_1^0$ in top pair, $1 \, {}^1 \!
  S_0^0$ in bottom pair) as a function of cutoff $\Lambda_\mu$ for
  $\Lambda_e^2=\Lambda_\mu^2 + 4(m_\mu^2 - m_e^2)$,
  $\alpha=0.3$, $m_e=\frac{1}{2}m_\mu$, $N=25$.  $\Lambda_\mu$ is
  given in units of the muon Bohr momentum $\alpha m_\mu/2$.  The
  ($\circ$) points indicate the full result precisely following the
  methods of Ref.~\cite{Trittmann:1997xz}, and the $(\mbox{\tiny
    $\blacksquare$} )$ points indicate the result after the
  implementation of a subtraction (described in the text) of the
  amplitude responsible for poor ultraviolet behavior in ${}^1S_0$
  channels.}
\end{center}
\end{figure}

\begin{figure}
\begin{center}
\includegraphics[width=\linewidth]{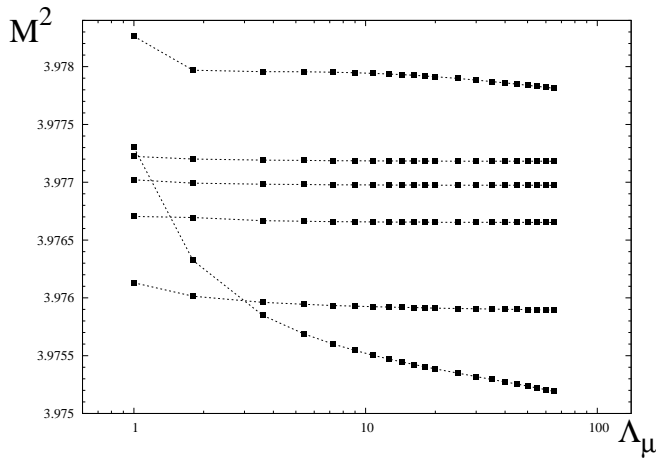}
\caption{\label{fig:cut2}Mass eigenvalues of $n=2$ true muonium states
  with $J_z=0$ (top to bottom: $2 \, {}^3 \! S_1^0$, $2 \, {} ^3 \!
  P_2^0$, $2 \, {}^1 \! P_1^0$, $2 \, {}^3 \! P_1^0$, $2 \, {}^3 \!
  P_0^0$, $2 \, {}^1 \! S_0^0$) as a function of cutoff $\Lambda_\mu$.
  The numerical inputs and units of $\Lambda_{e,\mu}$ are the same as
  in Fig.~\ref{fig:cut1}.  The amplitude subtraction described in the
  text has been performed for all states here.}
\end{center}
\end{figure}

To investigate the effect of this choice of cutoffs, we fix
$\Lambda_e$ as described above
and vary $1 \le \Lambda_\mu \le 65$ (in units of muon Bohr momentum
$\alpha m_\mu/2$).  Results for the $n = 1, 2$ eigenstates with
$m_e=\frac 1 2 m_\mu$ can be seen in Figs.~\ref{fig:cut1} and
\ref{fig:cut2}, respectively, which can be compared to Figs.~2.8 and
2.9 of Ref.~\cite{Trittmann:1997xz}, where the same study was
performed for positronium.  For an exact analogue to the results of
Ref.~\cite{Trittmann:1997xz}, one should compare the lines in
Fig.~\ref{fig:cut1} with open-circle points directly to their
analogues in the earlier work.  Since rotational invariance is
obscured in the front form, states studied in the numerical
simulations are labeled by adding the $J_z$ label, as in $n \,
{}^{2S+1} \! L_J^{J_z}$.

%

The most striking feature of the initial (open-circle points) results
in Fig.~\ref{fig:cut1} is the strong dependence on the cutoff
$\Lambda_\mu$ of the $1{}^1S_0$ mass eigenvalue compared to that of
$1{}^3S_1$ (A similar effect occurs for the $2{}^1S_0$ mass
eigenvalue).  One may initially wonder whether this effect is due to
an inappropriate handling of lepton mass renormalization.  The full
shift of the bound-state mass due to one-loop lepton mass
renormalization in front form is given by~\cite{Krautgartner:1991xz}
\begin{eqnarray}
\Delta M^2 & = &
\frac{\alpha}{2\pi} m^2 \left[ 3 \ln \left(
\frac{\Lambda^2 + m^2}{m^2} \right) - \frac{\Lambda^2}{\Lambda^2 +
m^2} \right] \nonumber \\ & &
\times \left( \frac{1}{x} + \frac{1}{1-x} \right) \, .
\label{eq:massrenorm}
\end{eqnarray}
This expression is obtained from the sum of loop and contraction
diagrams; as is well known ({\it e.g.}, see~\cite{Brodsky:1997de}),
the individual loop diagrams that give the renormalization constants
$Z_2$ or $Z_1$ in light-front form carry momentum dependence, but the
Ward identity guarantees that their sum does not, allowing one to
adopt an on-shell renormalization scheme in which the input $\alpha$
and $m$ values are given by the physical ones.  Since the higher-order
corrections not given here must necessarily subtract the $\ln \Lambda$
divergence of Eq.~(\ref{eq:massrenorm}) but generally produce
additional corrections, one may choose to remove the divergence in a
variety of ways.  The work of
Refs.~\cite{Krautgartner:1991xz,Trittmann:1997xz} advocates simply
taking $\Delta M^2 = 0$.  To gauge the effect of other choices, we
consider either subtracting from the bracketed term of
Eq.~(\ref{eq:massrenorm}) only the $\ln (\Lambda^2/m^2)$ portion, or
the $O[(m^2)^0]$ correction as well.  In the latter case, the
bound-state eigenvalues $M^2$ change by less than 1 part in $10^4$ by
the time $\Lambda_\mu$ is as small as $2m_\mu$.  We therefore also opt
for the simple choice $\Delta M^2 = 0$ in this work.

We also note that the derivation of Eq.~(\ref{eq:massrenorm}) neglects
one diagram, in which the leptons exchange an ``instantaneous'' photon
in the presence of a spectator photon, because it is non-diagonal in
the single-lepton spins and momenta, and therefore gives rise to a
self-mass correction of the atom that is not just a single-lepton mass
renormalization.  Certainly, such an effect could be included as an
$O(\alpha^2)$ perturbative correction.

Indeed, it is hard to imagine how simple lepton mass renormalization
would treat the ${}^1S_0$ states so differently from the others.
This phenomenon was noted as early as Ref.~\cite{Krautgartner:1991xz}.
Since the subsequent work of Ref.~\cite{Trittmann:1997xz} improved the
numerical quality of the $C=-1$ ${}^3S_1$ states by the inclusion of
the Fock state $\left|\gamma\right>$, one may expect the inclusion of
$\left|\gamma\gamma \right>$ to improve the $C=+1$ ${}^1S_0$ states.
We will present analysis of this effect in the future~\cite{LamLeb2},
but for the present adapt a regularization first suggested in
Ref.~\cite{Krautgartner:1991xz}: The strong dependence of ${}^1S_0$
states on $\Lambda$ can be traced to a portion of the matrix element
between lepton antiparallel-helicity states (called $G_2$ in App.~F.3
of Ref.~\cite{Trittmann:1997xz}) that approaches a constant as
$k_\perp \equiv |\bm{k}_\perp|$ or $k'_\perp \equiv |\bm{k}'_\perp|\to
\infty$ and therefore produces $\delta$ function-like behavior in
configuration space.  Including such dependence imposes a singular
ultraviolet behavior on the system, and therefore it must be
regularized; Ref.~\cite{Krautgartner:1991xz} chose simply to delete
this term from their calculations.  However, we choose to subtract
only its limit as $k_\perp$ or $k'_\perp \to \infty$, which retains
part of the term (including $x$ and $x'$ dependence).  Again, this
subtraction scheme is designed only to remove strong $\Lambda$
dependence in ${}^1S_0$ states that the explicit inclusion of the
$\left|\gamma\gamma \right>$ state must eventually address.  The
effect of this subtraction is shown as lines in
Figs.~\ref{fig:cut1},\ref{fig:cut2} with filled square points, and
demonstrates a great improvement in the stabilization of the $\Lambda$
dependence of ${}^1S_0$ states, with fairly minimal changes to that of
other states.  The subtraction for the $2 {}^1S_0$ state is not shown
in Fig.~\ref{fig:cut2}, but it amounts to a decrease in the
$\Lambda_\mu$ dependence by over a factor of 10.

\begin{figure}
\begin{center}
\includegraphics[width=\linewidth]{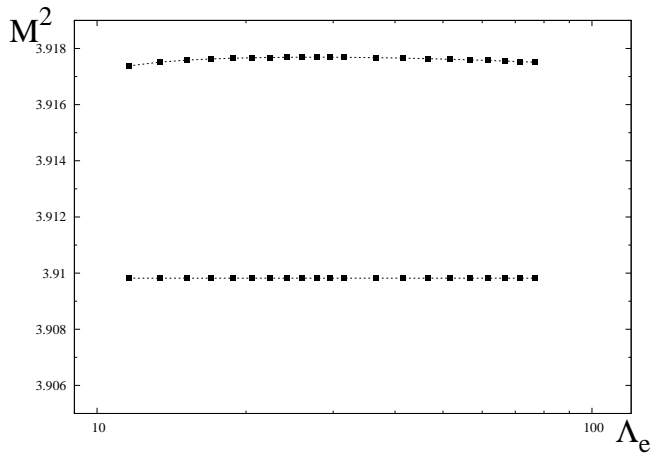}

\caption{\label{fig:mle}Mass eigenvalues of $n=1$ true muonium states
  with $J_z=0$ ($1 \, {}^3 \! S_1^0$ at top, $1 \, {}^1 \! S_0^0$ at
  bottom) as a function of cutoff $\Lambda_e$ in units of the muon
  Bohr momentum $\alpha m_\mu/2$, with $\Lambda_{\mu}=1$ in these
  units, $\alpha=0.3$, $m_e=\frac{1}{2}m_\mu$, $N=25$.}
\end{center}
\end{figure}

We explore the sensitivity of the results to varying $\Lambda_e$ in
Fig.~\ref{fig:mle}, where the mass eigenvalues of the $1{}^1S_0$ and
$1{}^3S_1$ states are reported as functions of $\Lambda_e$ in units of
the muon Bohr momentum $\alpha m_\mu/2$.  That is, $\Lambda_e \simeq
11.6$ corresponds to the value used in
Figs.~\ref{fig:cut1},\ref{fig:cut2}.  The results are quite
insensitive to larger $\Lambda_e$, which means that allowing for
greater off-shell momentum in the $\left|\ee\right>$ sector than in
our simple prescription has very little impact on the true muonium
spectrum.  On the other hand, decreasing $\Lambda_e^2$ below
$\Lambda_\mu^2 + 4(m_\mu^2 - m_e^2)$ (not depicted here) reveals
strong fluctuations in the dependence of mass eigenvalues upon
$\Lambda_e$, which we attribute to an undersampling of the continuum
$\left|\ee\right>$ states with invariant mass below the full true
muonium bound-state mass; the analogous effect in a hadronic physics
calculation would be interpreted as a failure of quark-hadron duality
due to insufficient phase space allowed for lighter components of the
hadron bound state.

\section{Results} \label{sec:results}

\begin{figure}
\begin{center}
\includegraphics[width=\linewidth]{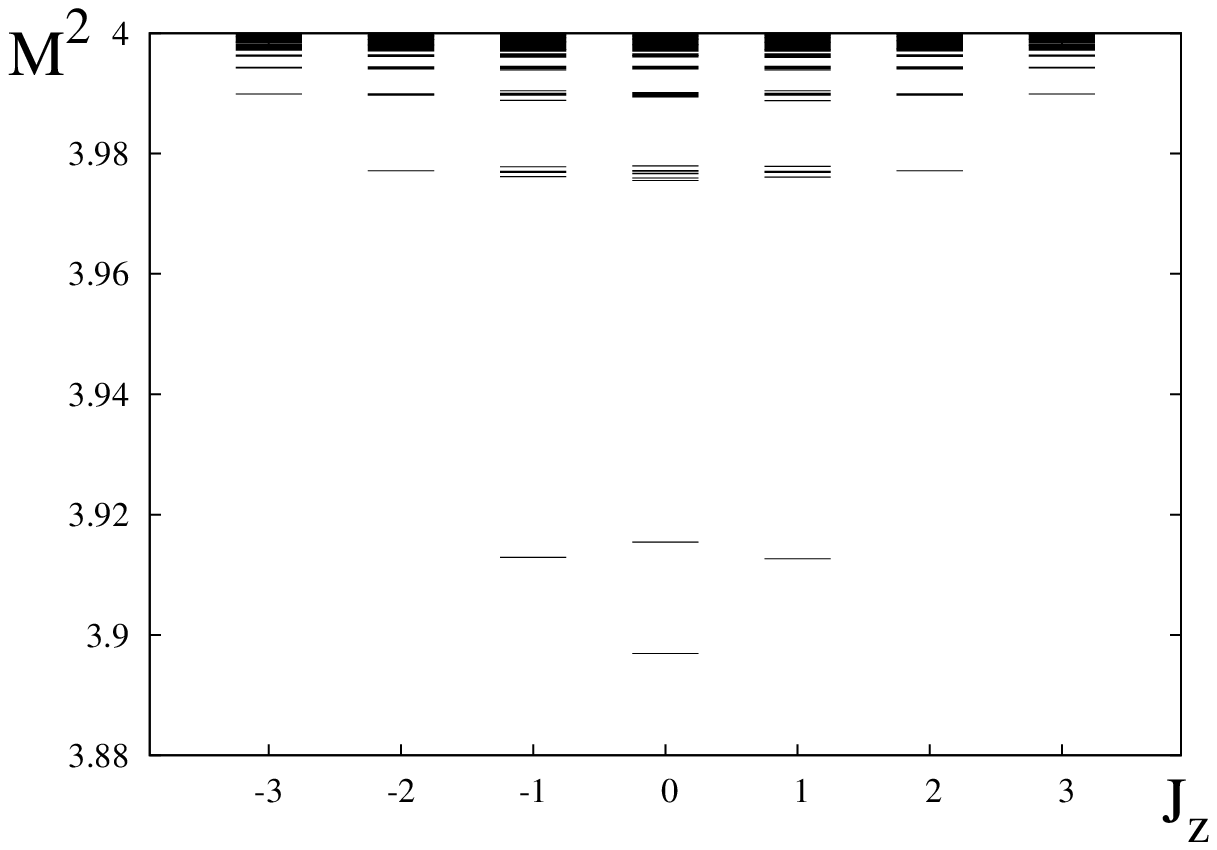}
\includegraphics[width=\linewidth]{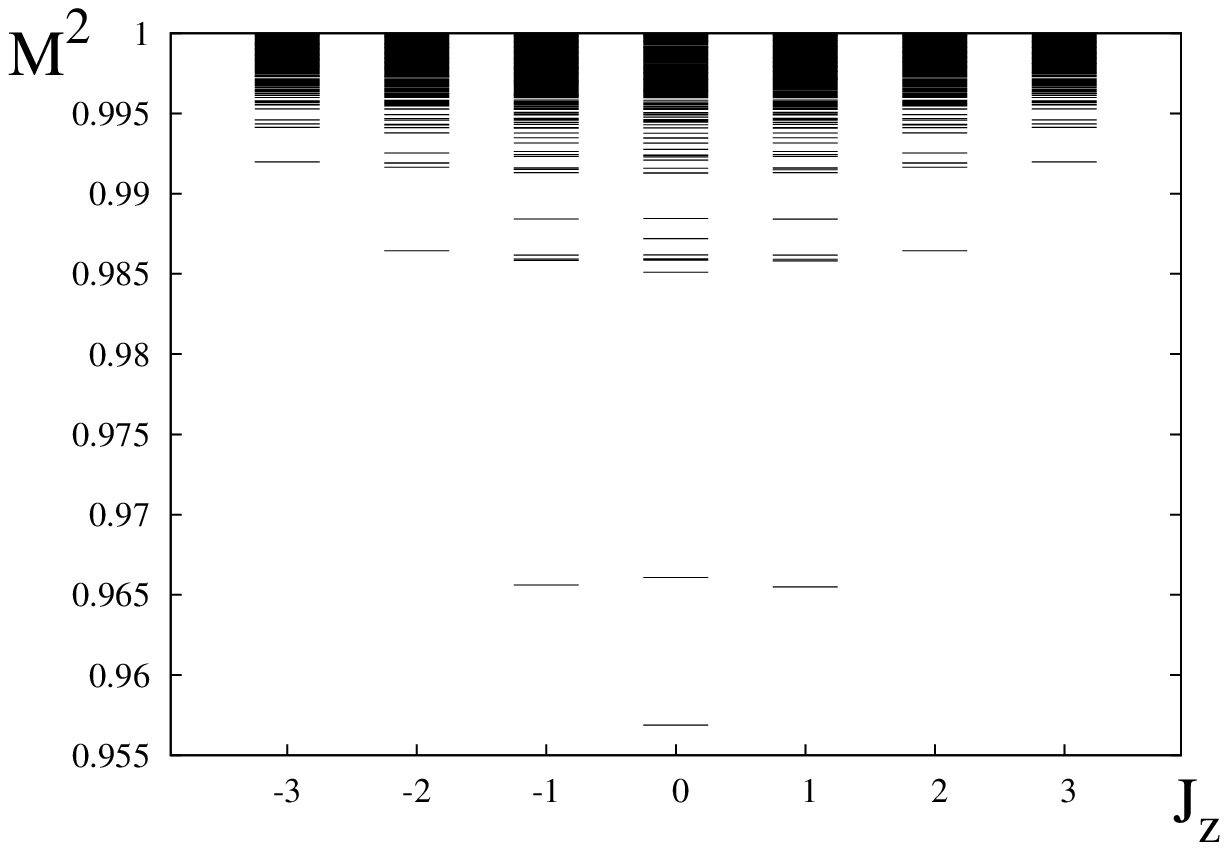}
\caption{\label{fig:elev}Spectrum of (top) true muonium and (bottom)
  positronium with $J_z=-3,-2, \ldots ,+3$.  The spectra are
  calculated using $\alpha=0.3$, $m_e=\frac{1}{2}m_\mu$,
  $\Lambda_\mu=10\alpha m_\mu/2$, $\Lambda_e=[\Lambda_\mu^2 +
  4(m_\mu^2 - m_e^2)]^{1/2} \simeq 15.3 \alpha m_\mu/2$,
  $N=25$.  The mass-squared eigenvalues $M^2_n$ are expressed in units
  of $m_\mu^2$.}
\end{center}
\end{figure}

Using a version of the code in Ref.~\cite{Trittmann:1997xz} modified
as discussed above, we compute the entire bound-state spectrum of true
muonium and positronium including valence Fock states of both
$|\mm\rangle$ and $|\ee\rangle$ for $J_z=-3,-2, \ldots ,+3$ ({\it
  e.g.}, Fig.~\ref{fig:elev}), taking $\alpha = 0.3$, $m_e = \frac 1 2
m_\mu$, $\Lambda_\mu = 10 \alpha m_\mu /2 \simeq 1.5 m_\mu$, and
$\Lambda_e = [\Lambda_\mu^2 + 4(m_\mu^2 - m_e^2)]^{1/2} \simeq 15.3
\alpha m_\mu /2 \simeq 2.3 m_\mu$.

From Fig.~\ref{fig:elev}, we see that the true muonium spectrum is
nearly identical to that found in \cite{Trittmann:1997xz}.  The shifts
caused by the inclusion of the $\ee$ sector are smaller than can be
resolved in this plot.  Likewise, the positronium spectrum indicates
multiplets with the expected multiplicities and ordering.

\begin{figure}
\begin{center}
\includegraphics[width=\linewidth]{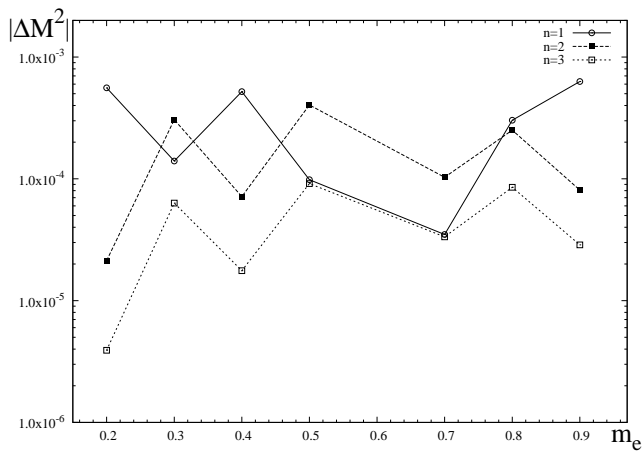}
\caption{
  \label{fig:mmj0}Eigenvalue shifts $\Delta M^2 \equiv
  M^2_{\mu\mu}-M^2_{0}$ (in units of $m_\mu^2$) for the $n \leq 3$,
  $J_z=0$ triplet states of true muonium as functions of $m_e$, for
  $\alpha=0.3$, $\Lambda_\mu=\alpha m_\mu/2$,
  $\Lambda_e=[\Lambda_\mu^2 + 4(m_\mu^2 - m_e^2)]^{1/2} \simeq 11.6
  \alpha m_\mu/2$, and values of $N$ are adjusted as described in the
  text.  From top to bottom, the states are $1 \, {}^3 \!  S^0_1$, $2
  \, {}^3 \! S^0_1$, and $3 \, {}^3 \! S^0_1$.}
\end{center}
\end{figure}

True muonium presents an extremely intriguing physical situation not
typically encountered in light-front studies, and particularly not in
light-front positronium studies: In the invariant mass range $4m_e^2 <
M^2 < 4m_\mu^2$, the $|\mm\rangle$ component is bound but the
$|\ee\rangle$ component forms a continuum.  The invariant mass $M_S$
of the $|\ee\rangle$ state satisfies the constraint
\begin{equation} \label{eq:contmass}
 M^2_S=\frac{m_e^2+\bm{k}_\perp^2}{x(1-x)} \, ,
\end{equation}
but is otherwise unconstrained.  Representing such states in the DLCQ
formulation presents interesting numerical challenges, analogous to
representing band structures in solid-state systems by closely-spaced
discrete energy levels.  Even so, since the two flavor sectors can
only interact through the single-photon annihilation channel, the only
true muonium states in this model affected by the inclusion of
$|\ee\rangle$ are those with $|J_z|\leq1$~\cite{Trittmann:1997xz}.
Denoting the bound-state mass-squared eigenvalue before and after
including the $\ee$ states as $M^2_{0}$ and $M^2_{\mu\mu}$,
respectively, Fig.~\ref{fig:mmj0} plots the magnitudes of the mass
shifts $\Delta M^2 \equiv M^2_{\mu\mu}-M^2_{0}$ of $J_z=0$ true
muonium states as a function of $m_e$ in the $n=1,2,3$ energy levels.

%

Although the results may seem noisy in $m_e$, one must first note that
the shifts $\Delta M^2$ are so small that they at no point lead to a
level crossing, and moreover, a trend is clearly visible that suggests
the shifts decrease quickly with increasing principal quantum number
$n$ (approximately as $1/n^3$, see below).  For the $J_z=0$ case, $n
\, {}^3 \! S_1^{\, 0}$ are the only states affected by the new sector
in a numerically significant way, in agreement with front-form
predictions~\cite{Trittmann:1997xz}.  One finds in the $|J_z|=1$ cases
(not plotted here) the $P$ states are also affected, but at a much
lower level,
and that the mass shifts for states differing only in $J_z$ are not
the same, reflecting that rotational invariance in the light-front
calculation at finite numerical accuracy is not entirely restored.


The reason for the fluctuations in Fig.~\ref{fig:mmj0} is just as
interesting as the results themselves.  As indicated above, the
$|\ee\rangle$ continuum states near the $|\mm\rangle$ bound states,
(which lie just below $M^2 = 4m_\mu^2$) are simulated numerically in
DLCQ as clusters of discrete energy levels rather than a true
continuum.  The location in $M^2$ of these clusters is determined by
$\mu$ [Eq.~(\ref{eq:mudef})], with the number of such clusters and the
size of gaps between them determined largely by $N_\mu$.  The density
of energy levels within each cluster is determined by $N_\theta$, but
we also note that the spacing of the levels within each cluster is not
entirely uniform, being more dense at larger values of $M^2$.

One expects that simply increasing the values of $N_\mu$ and
$N_\theta$ in the simulations must eventually suppress the numerical
artifacts associated with the discretization.  However, for the
moderate values ($N_\mu$, $N_\theta < 50$) used here, several features
make the analysis more complicated: First, the larger value of the
cutoff $\Lambda_e$ compared to $\Lambda_\mu$ (see
Sec.~\ref{sec:cutoffs}) allows for a substantial phase space to be
available to the $|\ee\rangle$ continuum states, only some of which
overlap with the $|\mm\rangle$ bound states, and this issue is
exacerbated as $\Lambda_e$ increases; in other words, only some of the
$|\ee\rangle$ clusters overlap with the $|\mm\rangle$ states, and
simply increasing $N_\mu$ does not directly alleviate this fact.
Related to this point is the nonlinear nature of the mapping
Eq.~(\ref{eq:weight}), which was designed to guarantee a sufficient
sampling of points up to $\mu = \frac{\Lambda_e}{2}$, but does not
necessarily suitably sample the region near the $|\mm\rangle$ bound
states.

The discrete sampling of these continuum states has a noticeable
effect on the shifts $\Delta M^2$.  Quite generally, we find that a
given $|\mm\rangle$ bound state prior to the inclusion of electrons
undergoes a shift in $\Delta M^2$ toward the energy levels of the
$|\ee\rangle$ states in the nearest clusters.  Clearly, such an effect
is a numerical artifact since the true $|\ee\rangle$ spectrum is
continuous, and the shift can be pronounced if the numerical
simulation is such as to produce no cluster of $|\ee\rangle$ states
near the original $|\mm\rangle$ bound state.  We report only results
from simulations in which the $|\mm\rangle$ state lies within an
$|\ee\rangle$ cluster; guaranteeing that this scenario occurs requires
a delicate balancing of the parameters $m_e$, $\Lambda_e$, $N_\mu$,
and attention to the nature of the mapping function $f(\mu)$.  Even in
the case that a $|\mm\rangle$ state lies neatly within an
$|\ee\rangle$ cluster, one must note that not every $|\ee\rangle$
state has the same quantum numbers as the $|\mm\rangle$ state and can
mix with it.  All of these effects must be taken into account in
understanding the nature of results like Fig.~\ref{fig:mmj0};
nevertheless, the fact remains that broad trends of definite physical
significance can still be identified.

For example, the addition of the $|\ee\rangle$ component should lead
to a modification of the Lamb shift (by which we mean the sum of all
radiative corrections) proportional to a power of the principal
quantum number $n$.  While the specific quantitative values of these
shifts show some sensitivity to the inputs, one might expect their
ratios for different states for any given set of simulation parameters
$m_e$, $\Lambda_e$, and $N$ to be less sensitive.  To study the Lamb
shift modifications, we take the ratio of the mass shifts for
different $n$ for $^3 \! S_1$ states.
We define the ratio $r_{nn'}$ for $n' > n$ via
\begin{equation}
 r_{nn'} \equiv \frac{\Delta M^2_n}{\Delta M^2_{n'}} \, .
\end{equation}
Taking the average of $r_{nn'}$ over all $m_e$ values used for the
computations, we determine the leading-order dependence $\Delta
M_n^2\propto n^{-\beta}$ from the relation
\begin{equation}
 \ln(r_{nn'})=-\beta\ln\left(\frac{n}{n'}\right) \, .
\label{eq:betadef}
\end{equation}
The results are presented in Table~\ref{tab:beta}.  We find that
$\beta\approx3$ for ${}^3 \! S_1$ states, which agrees with
instant-form perturbation theory calculations of Lamb
shifts~\cite{Itzykson:1980rh}.

\begin{table}[ht]
\begin{tabular}{l|ccc}
\hline\hline
$^{2S+1} \! L_{J}^{J_z}$ &\multicolumn{3}{|c}{$\beta$}\\
\hline
\multicolumn{1}{c|}{$n, \, n'$} & 1, 2 & 1, 3 & 2, 3 \\
\hline
$^3 \! S_1^0$ &$2.97\pm0.09$ &$3.3\pm0.3$&$3.3\pm0.3$\\
$^3 \! S_1^{-1}$ &$3.2\pm0.2$ &$3.3\pm0.2$&$3.6\pm0.4$\\
\hline
\end{tabular}
\caption{The exponent $\beta$ defined in Eq.~(\ref{eq:betadef}) for
  different states over the range $0.1\leq m_e/m_\mu\leq0.9$.  Errors
  are estimated from the variation in $m_e$ and $N$.}
\label{tab:beta}
\end{table}

\begin{figure}
\begin{center}
\includegraphics[width=\linewidth]{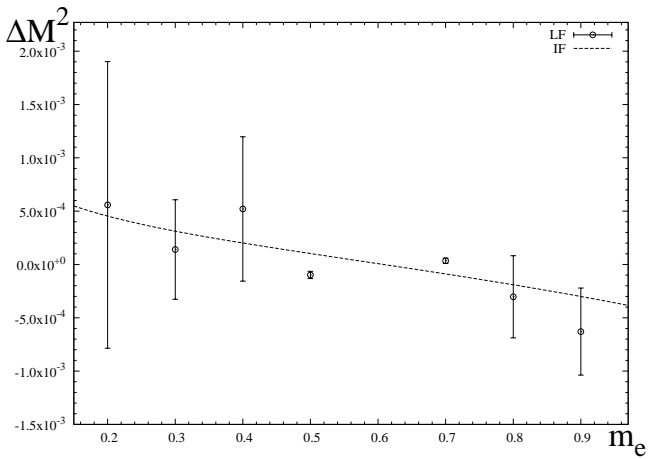}
\caption{
  \label{fig:mmet}Eigenvalue shifts $\Delta M^2 \equiv
  M^2_{\mu\mu}-M^2_{0}$ (in units of $m_\mu^2$) for $1 \, {}^3 \!
  S^0_1$.  The dashed line IF is the instant-form prediction from
  Eq.~(\ref{eq:dm}), using the nonrelativistic wave function, while
  the light-front (LF) points are obtained by taking $\alpha=0.3$,
  $\Lambda_\mu=\alpha m_\mu/2$, $\Lambda_e^2=[\Lambda_\mu^2 +
  4(m_\mu^2 - m_e^2)]^{1/2} \simeq 11.6 \alpha m_\mu/2$, and averaging
  over the results using several suitable values of $N$, as described
  in the text.}
\end{center}
\end{figure}

As discussed in Sec.~\ref{sec:ee}, one can compare the results of our
simulations to the predictions of nonrelativistic instant-form results
through Eqs.(\ref{eq:vp})--(\ref{eq:dm}).  Consider, for example,
$\Delta M^2$ of $1^3S_1^0$.  Even though the individual simulations at
particular fixed choices of $N_{\mu,\theta}$ for a given $m_e$ do not
rapidly converge to a single fixed value at the moderate values of
$N_{\mu,\theta}$ used here, if one restricts to simulations in which
the $|\mm\rangle$ state lies within a $|\ee\rangle$ cluster for the
given $m_e$, the eigenvalue shifts then lie in constrained ranges and
one may extract meaningful results by statistically averaging over the
results of these simulations, as exhibited in Fig.~\ref{fig:mmet}.
These light-front numerical results are seen in fact to agree fairly
well with the instant-form result, with a few important caveats:
First, the uncertainties become much larger for the smallest values of
$m_e$ (specifically seen in $m_e = 0.2 m_\mu$ in Fig.~\ref{fig:mmet}),
due to the increasing difficulty of properly sampling the
$|\ee\rangle$ clusters for small $m_e$ (on the other hand, simulations
using smaller values of $\alpha$ are not problematic, and simply serve
to decouple the true muonium and positronium spectra).  Second, the
tiny uncertainties at $m_e = 0.5 m_\mu$ and $0.7 m_\mu$ reflect the
accidental tendency of $|\ee\rangle$ clusters to appear in the region
of the $1 \, {}^3 \!  S^0_1$ $|\mm\rangle$ state.  Moreover, from the
formal point of view, the instant-form and light-front calculations
have three significant differences.
First, the instant-form result here represents only the real part of
the energy shift due to vacuum polarization and ignores, for example,
vertex corrections.  In front form, all of these effects are combined
together when one includes explicit $\left|\ee\right>$ and
$\left|\ee\gamma\right>$ states.  Second, vacuum polarization diagrams
in instant form contribute to the renormalization of the coupling
constant, an effect not taken into account in this simple model.
Finally, a result like Eq.~(\ref{eq:NREVP}) uses only the simplest
expression for the nonrelativistic wave function; instant-form
calculations that improve upon the nonrelativistic wave function
result appear in, {\it e.g.},
Refs.~\cite{Eiras:2000rh,Ivanov:2009zzd}.  Nevertheless, the level of
agreement in Fig.~\ref{fig:mmet} is gratifying.

\begin{figure*}
\begin{center}
\includegraphics[width=.27\linewidth,angle=-90]{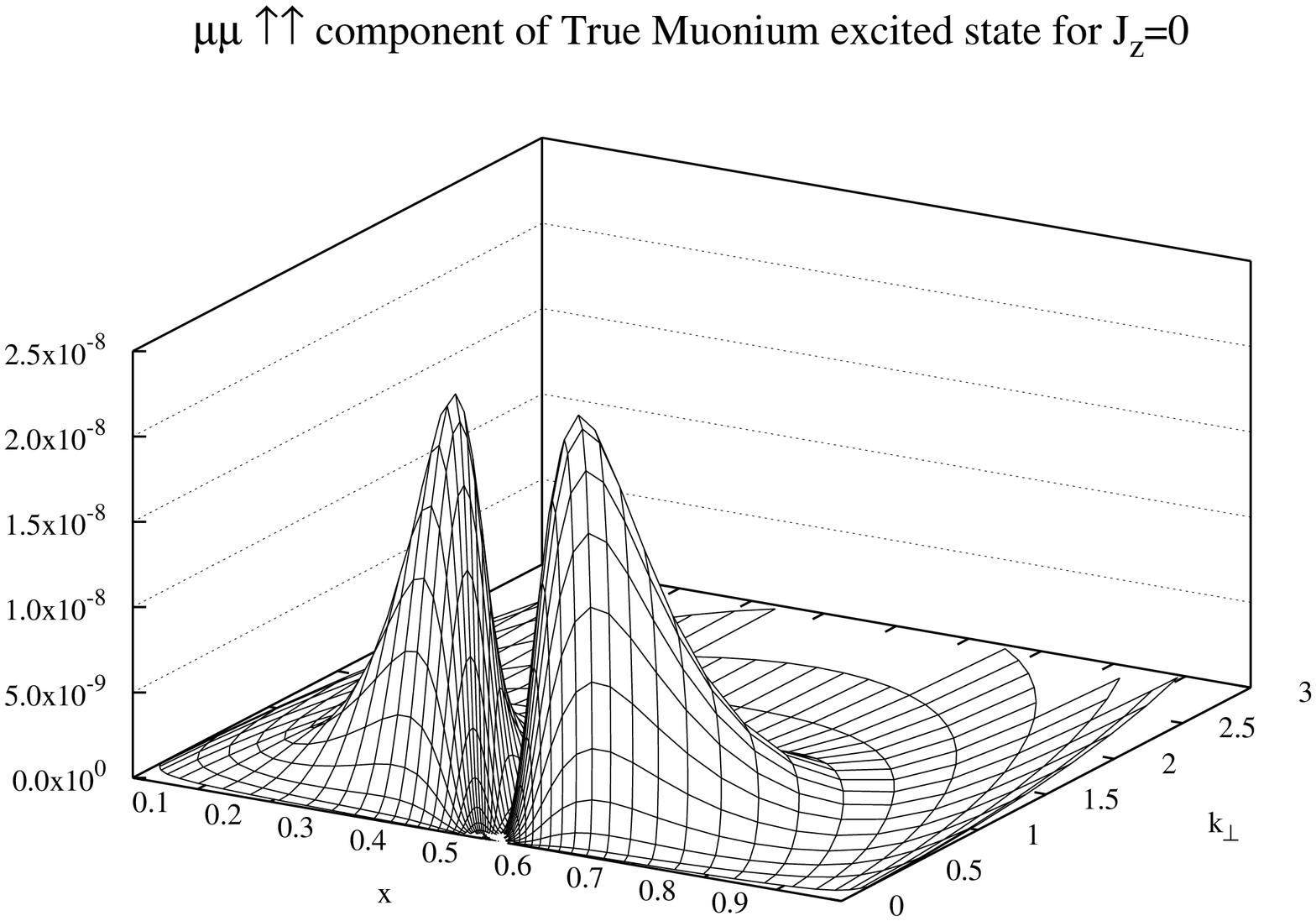}
\includegraphics[width=.27\linewidth,angle=-90]{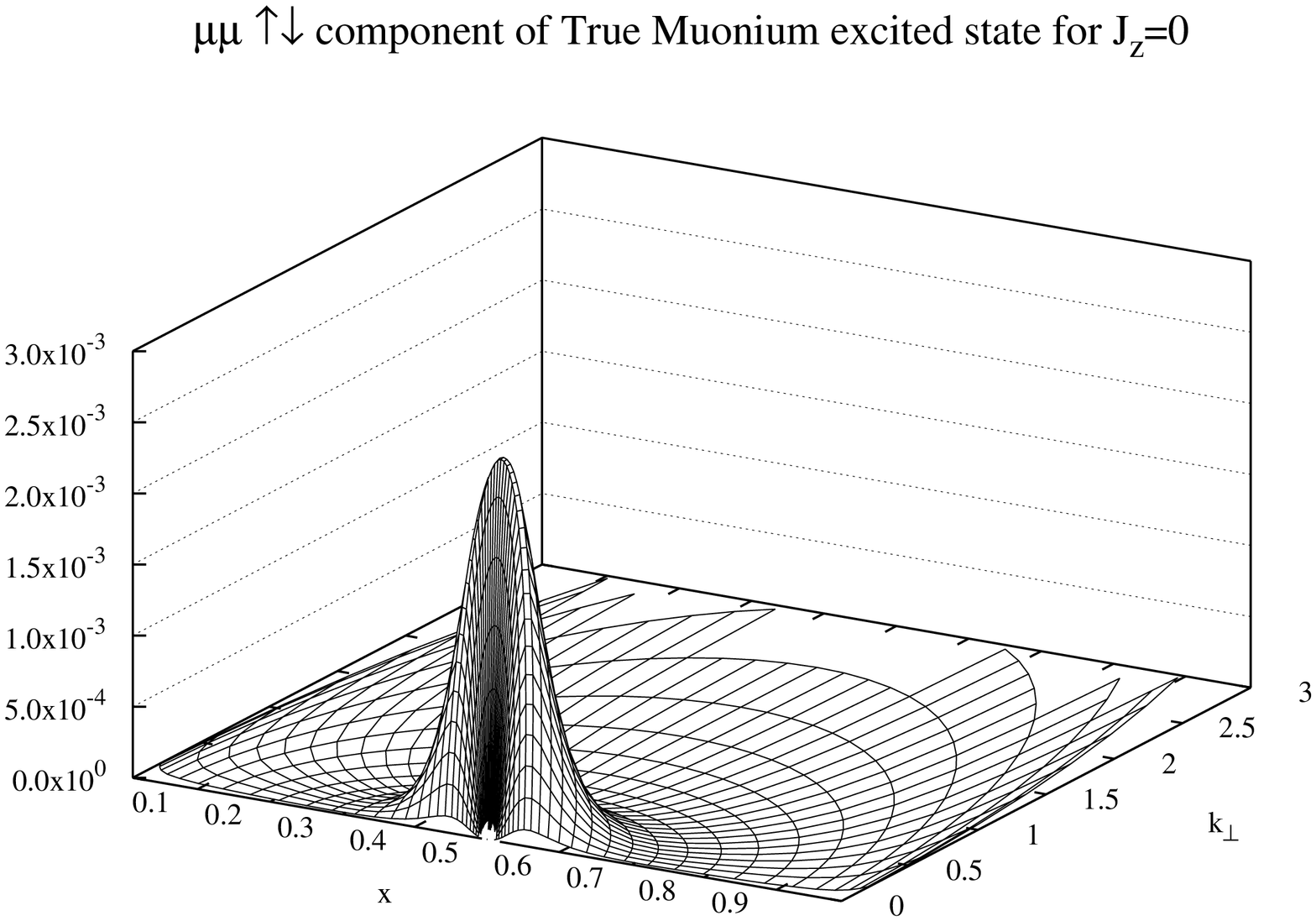}
\includegraphics[width=0.27\linewidth,angle=-90]{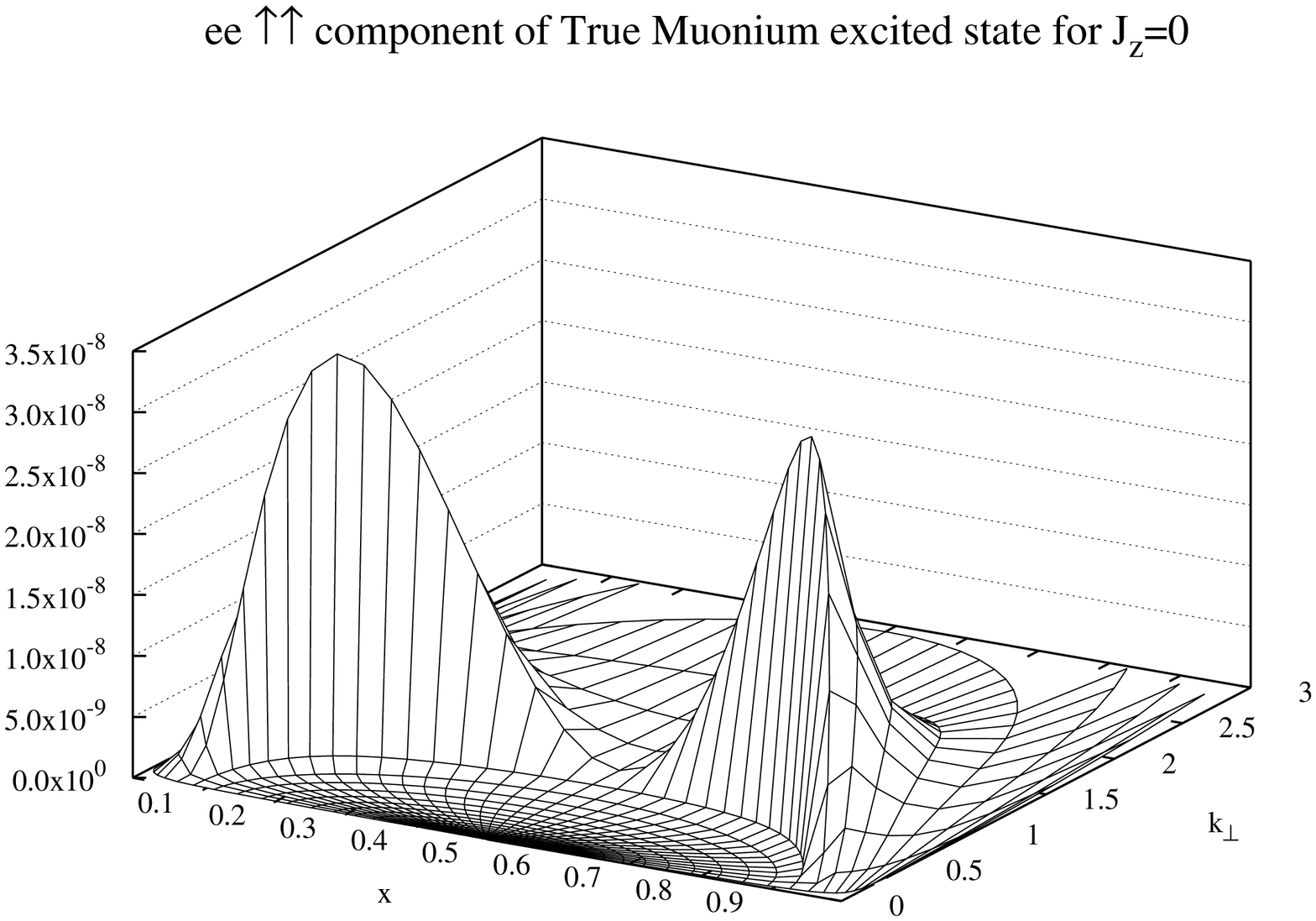}
\includegraphics[width=0.27\linewidth,angle=-90]{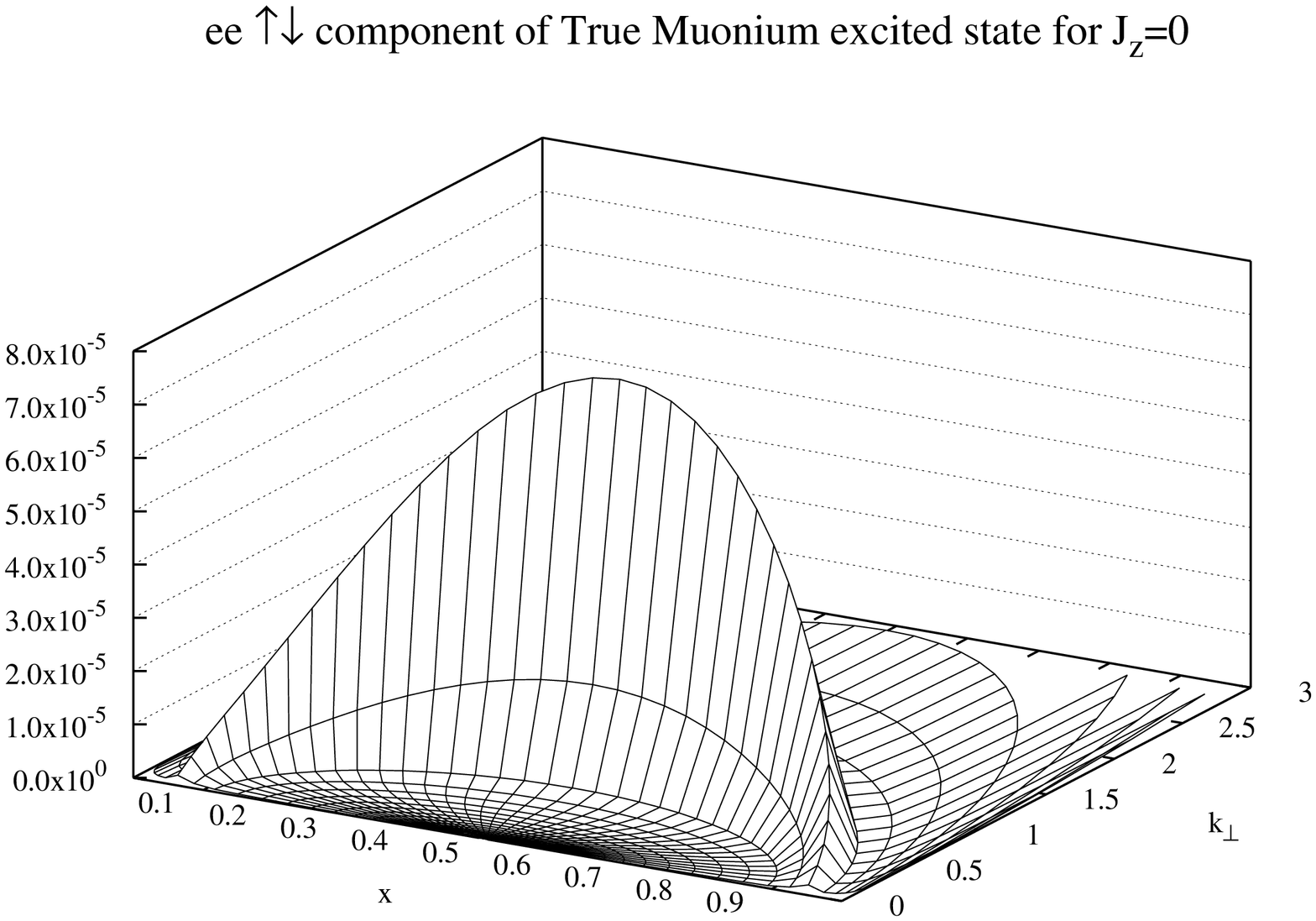}
\caption{\label{fig:mmw}The first triplet state ($1 \, {}^3 \! S_1^0$)
  probability density of the (top left) parallel-helicity muonic, (top
  right) antiparallel-helicity muonic, (bottom left) parallel-helicity
  electronic, and (bottom right) antiparallel-helicity electronic
  components of true muonium with $J_z=0$, as functions of $x$ and
  $k_\perp$, for $\alpha=0.3$, $m_e=\frac{1}{2}m_\mu$,
  $\Lambda_\mu=10\alpha m_\mu/2$, $\Lambda_e^2=\Lambda_\mu^2 +
  4(m_\mu^2 - m_e^2)$, and $N=37$.}
\end{center}
\end{figure*}

Inclusion of the $|\ee\rangle$ sector also changes the wave functions.
One expects that any true muonium bound state contains some component
of electron-positron continuum states.  The interaction of these
states with the true muonium should lead to a modification of the wave
functions by means of the $|\ee\rangle$ component.  As noted above,
the $1 \, {}^3 \!  S_1^0$ wave function is expected to be affected the
most by the inclusion of the $|\ee\rangle$.  In Fig.~\ref{fig:mmw} we
plot the probability density of the components of the $1 \, {}^3 \!
S_1^0$ state of true muonium using a particular set of parameters.
Noting the relative scales, one sees that the antiparallel components
dominates the state.  As expected for a bound state, the $\mm$ pair is
localized near $x=\frac 1 2$ and $k_\perp
\equiv |\bm{k}_\perp|=0$.

Again, the $|\ee\rangle$ wave function components of this state are
expected to appear as continuum states, since the invariant mass
$M^2>4m_e^2$.  From Fig.~\ref{fig:mmw}, one sees that the $e^+ e^-$
component is sharply peaked along the curve given by
Eq.~(\ref{eq:contmass}),
%
%
where $M^2_S$ is now also the invariant mass $M^2$ of the $1 \, {}^3
\! S_1^0$ state; the functional sharpness of such components suggests
how difficulties in numerical sampling can occur, as discussed in
Sec.~\ref{sec:cutoffs} and seen in Fig.~\ref{fig:mmj0}.  Additionally,
one sees that the fermionic nature of the electrons is manifested most
visibly in the parallel helicity component, where at $x=\frac 1 2$ the
probability is heavily suppressed.  One also sees that, in agreement
with the muonic components, the antiparallel component numerically
dominates.

Integrating over the coordinates $x,{k}_\perp$, we compile in
Table~\ref{tab:prob} the integrated probability for each component.
From this exercise, we find the surprising result that the
antiparallel electronic components contribute more to the wave
function than the parallel muonic component.  The large uncertainties
in the integrated probabilities for the continuum components arise
from their sensitive dependence upon $\Lambda_e$ and $N$ discussed in
Sec.~\ref{sec:cutoffs}.
\begin{table}[ht]
\begin{tabular}{l|c}
\hline\hline
Sector & $\int\mathrm{d}x \, \mathrm{d}^2 \! \bm{k}_\perp
P(x,{k}_\perp)$\\ \hline
$|\mu\bar{\mu},\uparrow\uparrow\rangle$&$(3.0\pm0.4)\times 10^{-6}$\\
$|\mu\bar{\mu},\uparrow\downarrow\rangle$&$0.49\pm0.01$\\
$|e\bar{e}, \, \uparrow\uparrow\rangle$&$\approx 10^{-7}-10^{-6}$\\
$|e\bar{e}, \, \uparrow\downarrow\rangle$&$\approx 10^{-3}-10^{-2}$\\
\hline
\end{tabular}
\caption{Integrated probability for each sector in the true muonium
  $1^3S_1^0$ state. Due to the parity invariance of QED, sectors with
  both helicities flipped have identical probability [{\it e.g.},
  $P(|\mu\bar{\mu},\uparrow\uparrow\rangle)=
  P(|\mu\bar{\mu},\downarrow\downarrow\rangle)]$, so the explicit
  numbers in this table should add to $\frac 1 2$.  The parameters
  used are $\alpha=0.3$, $m_e=\frac{1}{2}m_\mu$, $\Lambda_\mu=10\alpha
  m_\mu/2$, $\Lambda_e =[\Lambda_\mu^2 + 4(m_\mu^2 - m_e^2)]^{1/2}
  \simeq 15.3 \alpha m_\mu/2$, $N=37$.  Uncertainties are estimated by
  varying $N$.}
\label{tab:prob}
\end{table}

\section{Discussion and Conclusions} \label{sec:con}

In this paper, we have computed the light-front wave functions of true
muonium, using a simple model of two-flavor QED that includes a
limited number (5) of explicit Fock states.  Using this model, we have
seen that corrections from $|\ee\rangle$ have a noticeable effect on
true muonium states.  The probability density for the first excited
state of true muonium, the triplet $1 \, {}^3 \! S_1$ with $J_z = 0$,
was determined and the integrated probability for each valence
Fock-space component was computed.  If $m_e$ is taken to be a
substantial fraction of $m_\mu$ and the fine structure constant is
taken large ($\alpha = 0.3$), then the $|\ee\rangle$ states constitute
as much as $O(10^{-2})$ of the true muonium eigenstates.
 
For a more accurate model, three critical Fock states must also be
included: $|\gamma\gamma\rangle$, which dominates the decay of singlet
states of true muonium (and in particular should have a pronounced
effect on ${}^1 \! S_0$ wave functions), and the pair of states
$|\mm\ee\rangle$ and $|\mm\mm\rangle$, which provide crucial
contributions to the vacuum polarization corrections.  To integrate
these states into a model, a proper renormalization of the Hamiltonian
is necessary.  Methods of interest that have been developed to carry
out the renormalization include using Pauli-Villars
regulators~\cite{Chabysheva:2009vm,Chabysheva:2010vk}, the
Hamiltonian-flow approach~\cite{Gubankova:1998wj,Gubankova:1999cx},
methods with sector-dependent
counterterms~\cite{Karmanov:2008br,Karmanov:2012aj}, and the basis
light-front approach~\cite{Wiecki:2014ola}.  Moreover, a technique to
represent continuum $\left| \ee\right>$ states improved over the
discrete approach used here (for example, replacing them with an
equivalent spectral function~\cite{LamLeb2}) will produce numerically
more stable results.

In addition to the analytical work required, including these
additional Fock states will mandate a much larger numerical effort.
While the new states could in principle just be added to the list of
valence states included (just as $|\ee\rangle$ and $|\ee\gamma\rangle$
are this work), the numerical effort required to accommodate them is
likely too great.  For example, the sector $|\mm\ee\rangle$ has 6
independent coordinates and 16 spin states.  This sector alone would
naively require $16\times N^6$ elements in each eigenvector, a number
that is greater than the dimension of the entire Fock space considered
in our toy model.  Instead, the implementation of these Fock states
will likely require developing an appropriate effective interaction
for their successful incorporation into the calculation.

Once the component wave functions are computed by any suitable means,
they are ready to incorporate directly into the calculation of any
physical process.  For example, the dissociative diffraction of pions
in the presence of nuclear matter~\cite{Frankfurt:1993it} (used to
explore QCD color transparency~\cite{Aitala:2000hc}) employs
distribution functions that, in the light-front form, are none other
than the pion component wave functions~\cite{Ashery:2006zw}.
Essentially the same physics, but applied to the much simpler QED
case, appears in experiments discussed above~\cite{Banburski:2012tk}
that probe the dissociation of true muonium on high-$Z$ targets.

\appendix*
\section{Matrix Elements between Flavor Sectors} \label{sec:matrix}
The relevant matrix elements for the calculation can be obtained
through a straightforward generalization of the results in
Ref.~\cite{Trittmann:1997xz}, particularly Appendix~F\@.  Those
involving a single flavor ($\mu$ or $e$) are exactly the same as in
Appendices~F.1--F.3, once the appropriate fermion mass is used (not
counting the subtraction in amplitude $G_2$ of App.~F.3 described in
Sec.~\ref{sec:cutoffs}).  We also correct a typo in the expression for
amplitude $G_3$ in~\cite{Trittmann:1997xz}: $1-n \to 1+n$.  Only the
single-photon annihilation graphs (App.~F.4) that mix the flavors
differ.  Denoting the initial and final fermion masses as $m'$, $m$,
respectively (and similarly for $x$ and $k_\perp$), the key kinematic
quantity appearing in the propagators is the effective kinetic energy
\begin{equation}
 \omega^* \equiv \frac{1}{2}\left[\frac{m^2+\kp^2}{x(1-x)}
+\frac{m'^{2}+\kp'^{2}}{x'(1-x')}\right] \, ,
\end{equation}
which is a slightly different $\omega^*$ than is used for the standard
annihilation graphs in the absence of $\ee$.  This definition in
either case has the special significance in the method of iterated
resolvents, that it allows the truncation of neglected Fock states in
a controlled manner.  The amplitudes for the process then read:
\begin{align}
I_1(x,\kp;x',\kp') &= \frac{\alpha}{\pi}\frac{2m'm}{\omega^*}
\frac{1}{x(1-x)}\frac{1}{x'(1-x')}\delta_{|J_z|,1} \, ,
\end{align}
\begin{align}
I_2(x,\kp;x',\kp') &=\frac{\alpha}{\pi}\left[\frac{2}{\omega^*}
\frac{\kp\kp'}{xx'}\delta_{|J_z|,1}+4\delta_{J_z,0}\right] \, ,
\end{align}
\begin{align}
I_3(x,\kp;x',\kp') &= \frac{\alpha}{\pi}\frac{2m\lambda_1}{\omega^*}
\frac{1}{x(1-x)} \frac{\kp'}{1-x'}\delta_{|J_z|,1} \, ,
\end{align}
\begin{align}
I_4(x,\kp;x',\kp') &= -\frac{\alpha}{\pi}\left[\frac{2}{\omega^*}
\frac{\kp\kp'}{x'(1-x)}\delta_{|J_z|,1}-4\delta_{J_z,0}\right] \, .
\end{align}
Note that the only nonvanishing matrix elements have $|J_z|\leq 1$, a
restriction due to the angular momentum of the photon.  Abbreviating
$I_i (1,2) \equiv I_i (x,\kp;x',\kp')$, the amplitudes for all allowed
combinations of helicity states are given in Table~\ref{tab:hel}.
\begin{table}[h]
\centerline{
\begin{tabular}{|c||c|c|c|c|}\hline
\rule[-3mm]{0mm}{8mm}{\bf $m:m'$} & $\uparrow\uparrow$ 
& $\uparrow\downarrow$ 
& $\downarrow\uparrow$ &\phantom{xx}
$\downarrow\downarrow$\phantom{xx} \\ \hline\hline
\rule[-3mm]{0mm}{8mm}$\uparrow\uparrow$ & 
$I_1(1,2)$   
&$I_3(2,1)$ & $I^*_3(2,1)$ & $0$ \\ \hline
\rule[-3mm]{0mm}{8mm}$\uparrow\downarrow$ & 
$I_3(1,2)$ 
& $I^*_2(1,2)$ & $I_4(2,1)$ &$0$ \\ \hline
\rule[-3mm]{0mm}{8mm}$\downarrow\uparrow$& 
$I_3^*(1,2)$ & $I_4(1,2)$ & $I_2(1,2)$  & $0$\\ \hline
\rule[-3mm]{0mm}{8mm} $\downarrow\downarrow$ & $0$ 
& $0$ & $0$ & $0$  \\
\hline
\end{tabular}
}
\caption{\label{tab:hel}Helicity table 
  of the annihilation graph, where $m$ and $m'$ indicate the
  final-state and initial-state fermions, respectively.}
\end{table}

\begin{acknowledgments}
  We thank S.~Brodsky for very useful conversations, and H.L.~thanks
  U.~Trittmann for valuable insight into understanding his code.  This
  work was supported by the National Science Foundation under Grant
  No.\ PHY-1068286.
\end{acknowledgments}
 
\bibliographystyle{apsrev4-1}

\begin{thebibliography}{10}%

\bibitem{Coombes:1976hi} 
  R.~Coombes, R.~Flexer, A.~Hall, R.~Kennelly, J.~Kirkby, R.~Piccioni,
  D.~Porat and M.~Schwartz {\it et al.},
  Phys.\ Rev.\ Lett.\  {\bf 37}, 249 (1976).

\bibitem{Cassidy:2007}
  D.B.~Cassidy and A.P.~Mills,
  Nature (London) {\bf 449}, 195 (2007).
  
\bibitem{Brodsky:2009gx} 
  S.J.~Brodsky and R.F.~Lebed,
  Phys.\ Rev.\ Lett.\  {\bf 102}, 213401 (2009)
  [arXiv:0904.2225 [hep-ph]].

\bibitem{Banburski:2012tk} 
  A.~Banburski and P.~Schuster,
  Phys.\ Rev.\ D {\bf 86}, 093007 (2012)
  [arXiv:1206.3961 [hep-ph]].

\bibitem{Antognini:1900ns} 
  A.~Antognini, F.~Nez, K.~Schuhmann, F.D.~Amaro, F.~Biraben,
  J.M.R.~Cardoso, D.S.~Covita and A.~Dax {\it et al.},
  Science {\bf 339}, 417 (2013).

\bibitem{TuckerSmith:2010ra} 
  D.~Tucker-Smith and I.~Yavin,
  Phys.\ Rev.\ D {\bf 83}, 101702 (2011)
  [arXiv:1011.4922 [hep-ph]].

\bibitem{Robiscoe:1965}
  R.T.~Robiscoe,
  Phys.\ Rev.\ {\bf 138}, A22 (1965).

\bibitem{Dirac:1949cp} 
  P.A.M.~Dirac,
  Rev.\ Mod.\ Phys.\  {\bf 21}, 392 (1949).

\bibitem{Brodsky:1997de} 
  S.J.~Brodsky, H.-C.~Pauli and S.S.~Pinsky,
  Phys.\ Rept.\  {\bf 301}, 299 (1998)
  [hep-ph/9705477].

\bibitem{Pauli:1985ps} 
  H.C.~Pauli and S.J.~Brodsky,
  Phys.\ Rev.\ D {\bf 32}, 2001 (1985).

\bibitem{Tang:1991rc} 
  A.C.~Tang, S.J.~Brodsky and H.C.~Pauli,
  Phys.\ Rev.\ D {\bf 44}, 1842 (1991).

\bibitem{Krautgartner:1991xz} 
  M.~Krautg\"artner, H.C.~Pauli and F.~W\"olz,
  Phys.\ Rev.\ D {\bf 45}, 3755 (1992).

\bibitem{Kaluza:1991kx} 
  {M.~Kalu\v{z}a} and H.C.~Pauli,
  Phys.\ Rev.\ D {\bf 45}, 2968 (1992).

\bibitem{Trittmann:1997xz} 
  U.~Trittmann and H.-C.~Pauli,
  hep-th/9704215.

\bibitem{Brodsky:1973kb} 
  S.J.~Brodsky, R.~Roskies and R.~Suaya,
  Phys.\ Rev.\ D {\bf 8}, 4574 (1973).

\bibitem{LamLeb2}
  H.~Lamm and R.F.~Lebed, in preparation.

\bibitem{Lepage:1980fj} 
  G.P.~Lepage and S.J.~Brodsky,
  Phys.\ Rev.\ D {\bf 22}, 2157 (1980).

\bibitem{Tamm:1945qv} 
  I.~Tamm,
  J.\ Phys.\ (USSR) {\bf 9}, 449 (1945).

\bibitem{Dancoff:1950ud} 
  S.M.~Dancoff,
  Phys.\ Rev.\  {\bf 78}, 382 (1950).

\bibitem{Pauli:1996dm} 
  H.-C.~Pauli,
  hep-th/9608035.

\bibitem{Trittmann:1997up} 
  U.~Trittmann,
  in {\it Les Houches 1997, New non-perturba\-tive methods and
    quantization on the light cone}, pp.\ 89-96
  [hep-th/9706055].

\bibitem{Trittmann:1997tt} 
  U.~Trittmann,
  hep-th/9705072.

\bibitem{Trittmann:2000gk} 
  U.~Trittmann and H.-C.~Pauli,
  Nucl.\ Phys.\ Proc.\ Suppl.\  {\bf 90}, 161 (2000).

\bibitem{Pauli:1997ns} 
  H.-C.~Pauli,
  in {\it Nagoya 1996, Perspectives of strong coupling gauge
    theories}, pp.\ 342--352 [hep-th/9706036].

\bibitem{Jentschura:1997tv} 
  U.D.~Jentschura, G.~Soff, V.G.~Ivanov and S.G.~Karshenboim,
  Phys.\ Rev.\ A {\bf 56}, 4483 (1997)
  [physics/9706026].
  
\bibitem{Jones:1996vy} 
  B.D.~Jones, R.J.~Perry and S.D.~G{\l}azek,
  Phys.\ Rev.\ D {\bf 55}, 6561 (1997)
  [hep-th/9605231].

\bibitem{Jones:1997cb} 
  B.D.~Jones,
  hep-th/9703106.

\bibitem{Karplus:1952wp} 
  R.~Karplus and A.~Klein,
  Phys.\ Rev.\  {\bf 87}, 848 (1952).

\bibitem{Chabysheva:2009vm} 
  S.S.~Chabysheva and J.R.~Hiller,
  Phys.\ Rev.\ D {\bf 81}, 074030 (2010)
  [arXiv:0911.4455 [hep-ph]].

\bibitem{Chabysheva:2010vk} 
  S.S.~Chabysheva and J.R.~Hiller,
  Phys.\ Rev.\ D {\bf 82}, 034004 (2010)
  [arXiv:1006.1077 [hep-ph]].

\bibitem{Gubankova:1998wj} 
  E.L.~Gubankova, H.-C.~Pauli and F.J.~Wegner,
  hep-th/9809143.

\bibitem{Gubankova:1999cx} 
  E.L.~Gubankova and G.~Papp,
  hep-th/9904081.

\bibitem{Karmanov:2008br} 
  V.A.~Karmanov, J.-F.~Mathiot and A.V.~Smirnov,
  Phys.\ Rev.\ D {\bf 77}, 085028 (2008)
  [arXiv:0801.4507 [hep-th]].

\bibitem{Karmanov:2012aj} 
  V.A.~Karmanov, J.F.~Mathiot and A.V.~Smirnov,
  Phys.\ Rev.\ D {\bf 86}, 085006 (2012)
  [arXiv:1204.3257 [hep-th]].

\bibitem{Wiecki:2014ola} 
  P.~Wiecki, Y.~Li, X.~Zhao, P.~Maris and J.P.~Vary,
  arXiv:1404.6234 [nucl-th].

\bibitem{Itzykson:1980rh} C.~Itzykson and J.-B.~Zuber, {\it Quantum
    Field Theory}, McGraw-Hill, New York (1980).

\bibitem{Eiras:2000rh} 
  D.~Eiras and J.~Soto,
  Phys.\ Lett.\ B {\bf 491}, 101 (2000)
  [hep-ph/0005066].
  
\bibitem{Ivanov:2009zzd} 
  V.G.~Ivanov, E.Y.~Korzinin and S.G.~Karshenboim,
  Phys.\ Rev.\ D {\bf 80}, 027702 (2009).
  
\bibitem{Frankfurt:1993it} 
  L.~Frankfurt, G.A.~Miller and M.~Strikman,
  Phys.\ Lett.\ B {\bf 304}, 1 (1993)
  [hep-ph/9305228].

\bibitem{Aitala:2000hc} 
  E.M.~Aitala {\it et al.}  [E791 Collaboration],
  Phys.\ Rev.\ Lett.\  {\bf 86}, 4773 (2001)
  [hep-ex/0010044].

\bibitem{Ashery:2006zw} 
  D.~Ashery,
  Prog.\ Part.\ Nucl.\ Phys.\  {\bf 56}, 279 (2006).

\end{thebibliography}

\end{document}